\documentclass[%
reprint,superscriptaddress, amsmath,amssymb,aps,prx, %linenumbers,
floatfix,]{revtex4-2}
\usepackage{graphicx}
\usepackage{dcolumn}
\usepackage{bm}
\usepackage{hyperref}
\hypersetup{colorlinks=true, citecolor=blue, urlcolor=blue, linkcolor=blue}
\usepackage{braket}
\usepackage{accents}
\usepackage{multirow,microtype}
\usepackage{dcolumn,booktabs}
\usepackage[version=4]{mhchem}
\usepackage{braket}

\begin{document}
	
\preprint{APS/123-QED}	

\title{Equivalence of charged and neutral density functional formulations for correcting the many-body self-interaction of polarons}
	
\author{Stefano Falletta}
\email{stefanofalletta@g.harvard.edu}
\affiliation{John A.\ Paulson School of Engineering and Applied Sciences, Harvard University, Cambridge, MA, USA}
\affiliation{Chaire de Simulation \`a l'Echelle Atomique (CSEA), Ecole Polytechnique F\'ed\'erale de Lausanne (EPFL), CH-1015 Lausanne, Switzerland}

\author{Jennifer Coulter}
\affiliation{John A.\ Paulson School of Engineering and Applied Sciences, Harvard University, Cambridge, MA, USA}

\author{Joel B. Varley}
\affiliation{Lawrence Livermore National Laboratory, Livermore, California, USA}
	
\author{Daniel \AA{}berg}
\affiliation{Lawrence Livermore National Laboratory, Livermore, California, USA}

\author{Babak Sadigh}
\affiliation{Lawrence Livermore National Laboratory, Livermore, California, USA}

\author{Boris Kozinsky}
\affiliation{John A.\ Paulson School of Engineering and Applied Sciences, Harvard University, Cambridge, MA, USA}
\affiliation{Robert Bosch LLC Research and Technology Center, Watertown, MA, USA}
		
\author{Alfredo Pasquarello}
\affiliation{Chaire de Simulation \`a l'Echelle Atomique (CSEA), Ecole Polytechnique F\'ed\'erale de Lausanne (EPFL), CH-1015 Lausanne, Switzerland}
	
\date{\today}
	
\begin{abstract} 
The electron self-interaction problem in density functional theory affects the accurate modeling of polarons, particularly their localization and formation energy. Charged and neutral density functional formulations have been developed to address this issue, yet their relationship remains unclear. Here, we demonstrate their equivalence in treating the many-body self-interaction of the polaron state. In particular, we connect with each other piecewise-linear functionals based on adding an extra charge to the supercell, the pSIC approach derived from the energetics of the neutral defect with polaronic distortions in a supercell, and the unit-cell method for polarons based on electron–phonon couplings. We show that that these approaches lead to the same formal expression of the self-interaction corrected energy, which is fully defined by the energetics of the neutral charge state of the charged polaronic structure. Residual differences between these methods solely arise from the achieved polaronic structure, which is affected by different treatments of electron-screening and finite-size effects. We apply these methods to a set of prototypical small hole and electron polarons, including the hole polaron in MgO, the hole polaron in $\beta$-Ga$_2$O$_3$, the $V_\text{k}$ center in NaI, the electron polaron in BiVO$_4$, and the electron polaron in TiO$_2$. We show that the ground-state properties of polarons obtained using charged and neutral density functional formulations are in excellent agreement. 
\end{abstract}
	
\maketitle

\section{Introduction}

% Electron self-interaction -> current challenge
The electron self-interaction is a spurious interaction that emerges in density functional theory (DFT) because of the underlying approximations in the exchange-correlation functional \cite{perdew1981PRB}. In addressing the electron self-interaction, it is important to distinguish between one-electron and many-electron systems. For a one-electron system, the self-interaction arises from the presence of Hartree and exchange-correlation potentials of the one-electron density, which are absent in the exact Schrödinger equation. For a many-electron system, one-body and many-body forms of self-interaction can be defined. The one-body self-interaction arises from the self-terms of the Coulomb interaction. At variance, the many-body self-interaction corresponds to the deviation of the total energy from being piecewise-linear upon electron addition \cite{perdew1982PRL,ruzsinszky2007JCP,zhang1998JPC,yang2000PRL,mori2006JCP}. In particular, the exact functional is piecewise linear and thus free from many-body self-interaction, as initially highlighted by Perdew \textit{et al.}\ \cite{perdew1982PRL} and Yang \textit{et al.}\ \cite{yang2000PRL}, with physical interpretations based on ensemble states and replicas, respectively. However, common DFT functionals exhibit a concavity with respect to the fractional number of electrons, thus deviating from the piecewise-linearity condition and lacking derivative discontinuities at integer electron occupations \cite{perdew1982PRL,kronik2020PCCP}. While the superiority of the many-body self-interaction over the one-body self-interaction has been established \cite{falletta2022PRL,falletta2022PRB}, a general understanding of the connections between various approaches introduced in the literature to address the self-interaction is still missing. 

% Importance of the many-body self-interaction: the case of the H dimer
The importance of addressing the many-body self-interaction can be easily shown even for a simple system such as the molecular hydrogen cation. In this case, suppressing the many-body self-interaction yields equilibrium bond length and dissociation energy in closer agreement with experiment \cite{ruzsinszky2007JCP}. Another significant aspect concerns the localization of charges in  Kohn-Sham DFT, which is often not achieved by standard semilocal functionals. Indeed, standard semilocal functionals favor charge delocalization due to the spurious inclusion of the electron self-interaction. For instance, in the case of the dissociation of the molecular hydrogen cation, a standard semilocal functional would favor the system with the electron half-split between the two protons, as opposed to the two degenerate configurations in which the electron is localized on either one of the protons.

% polarons
A striking example of charge delocalization in DFT is represented by polarons. Polarons are quasiparticles consisting of a localized charge coupled with self-induced lattice distortions \cite{franchini2021NRM}. In semiconductors or insulators, the formation of a polaron results in the introduction of an energy level within the band gap. Specifically, an electron polaron leads to an occupied energy level below the conduction band, whereas a hole polaron leads to an unoccupied energy level above the valence band. The stability of polarons in semiconductors is given by the balance between two contributions, namely the energy gain due to charge localization and the energy cost due to lattice distortions. The energy gain due to charge localization is related to the energy separation between the polaron level and the respective level of the delocalized band-edge state, and is reduced by the electron self-interaction. The energy cost due to lattice distortions is measured by the energy required to distort the lattice bonds from the pristine bulk structure to accommodate the polaron. When the energy cost due to lattice distortions overcomes the energy gain due to charge localization, the polaron state destabilizes. Consequently, the polaron charge delocalizes, and the polaronic lattice distortions vanish. The effects of self-interaction on charge delocalization are particularly crucial in the case of small polarons, which are localized over short length scales comparable to the lattice bonds. This can be quantitatively studied by using hybrid functionals \cite{perdew1996JCP} and admixing a fraction of Fock exchange to the semilocal exchange \cite{perdew1996PRL}. Polaron formation energies are found to heavily depend on the mixing fraction and thus on the description of the self-interaction \cite{kokott2018NJP,falletta2022PRL,falletta2022PRB,palermo2024PRB}.  Specifically, the Hartree-Fock functional leads to overestimated formation energies, as well as to largely overestimated band gaps. When the fraction of Fock exchange in hybrid functionals is chosen to enforce the piecewise linearity of the total energy, the polaron formation energies are robust  compared to other piecewise linear functionals \cite{falletta2022PRL,falletta2022PRB,falletta2022npj,falletta2023PRB,falletta2024JAP} and the band gaps are accurate when compared with experiment \cite{miceli2018PRB,deak2017PRB,bischoff2019PRM,bischoff2019PRB,bischoff2021PRR,wing2021PNAS,yang2022JPCL,falletta2022PRL,falletta2022PRB,yang2023npj,franckel2025PRB}.

% Analytical methods for correcting the self-interaction
Various approaches based on adding a self-interaction correction term to the energy functional have been introduced. In 1981, Perdew-Zunger proposed a scheme based on suppressing one-electron self-interaction contributions by subtracting Hartree and exchange-correlation potentials from the Kohn-Sham Hamiltonian \cite{perdew1981PRB}. Other approaches to suppress the one-electron self-interaction have been introduced, based on a mean field approximation \cite{ciofini2003CPL}, on the use of Fermi-Löwdin orbitals \cite{pederson2014JCP}, or on the use of orbitals determined from the density matrix \cite{paralta2024JPCA}. Alternatively, one can use the Hartree-Fock functional to exactly cancel the Hartree contributions generated by one-electron densities. Indeed, when applying the Fock exchange operator to a wave function state, one obtains a contribution that cancels the Hartree potential generated by that state. In this sense, the Hartree-Fock functional is free from one-body self-interaction. Next, d’Avezac \textit{et al.}\ introduced a modified version of the Perdew-Zunger correction to specifically address the self-interaction of the excess charge \cite{davezac2005PRB}. VandeVondele and Sprik proposed tunable versions of this category of functionals, which  further  improve the system energetics \cite{vandevondele2005PCCP}. Sadigh, Erhart, and \AA{}berg developed the pSIC method for localizing polarons through suppression of the many-body self-interaction, based on minimizing an energy functional defined as the sum of the energy of neutral system and its chemical potential \cite{sadigh2015PRB}. Sio \textit{et al}.\ introduced an \emph{ab initio} formulation of polarons based on electron-phonon couplings calculated for the unit cell \cite{sio2019PRL,sio2019PRB}, which is derived by removing self-interaction terms from a perturbative expansion of the total energy with respect to the polaron density. More recently, a unified hybrid-functional formulation for one-body and many-body self-interactions has been introduced \cite{falletta2022PRL,falletta2022PRB}. This formulation shows that, in the case of the PBE0($\alpha$) hybrid functional \cite{perdew1996JCP}, the many-body self-interaction energy correction can be expressed as the one-body self-interaction energy correction divided by the high-frequency dielectric constant. This shows that correcting for the many-body self-interaction allows one to account for electron screening effects, which are not considered when correcting for the one-body self-interaction \cite{falletta2022PRL,falletta2022PRB}. This demonstrates the superiority of the many-body form of self-interaction over the one-body form of self-interaction, at least for systems with more than one electron.

% nonempirical functional tuning
In addition to analytical methods, approaches based on functional tuning have been proposed to address the many-body self-interaction. Hybrid functionals have been extensively used to suppress the many-body self-interaction through the enforcement of the piecewise-linearity of the total energy \cite{varley2012PRB,elmaslmane2018PRM,carey2019JPCC,quirk2020JPCC,carey2021JPCC,carey2018JPCC,elmaslmane2018JCTC,miceli2018PRB,deak2017PRB,kronik2020PCCP,miceli2018PRB,sadigh2015PRB,sai2011PRL,refaely2013PRB,miceli2018PRB,deak2017PRB,bischoff2019PRM,bischoff2019PRB,osterbacka2020CM,bischoff2021PRR,yang2022JPCL,yang2023npj,kokott2018NJP,wing2020PRM,palermo2024PRB}. However, the computational cost of hybrid functional calculations is significantly higher than that of semilocal calculations, and becomes impractical for large systems or for \emph{ab initio} molecular dynamics. To overcome this limitation, Hubbard corrected DFT+$U$ functionals \cite{anisimov1991PRB,anisimov1991PRB2,anisimov1993PRB,solovyev1994PRB,czyifmmode1994PRB,liechtenstein1995PRB,anisimov1997JPCM,dudarev1998PRB,petukhov2003PRB,cococcioni2005PRB} can be used to suppress the many-body self-interaction by introducing on-site Coulomb repulsion for the occupation of the orbitals constituting the polaron state \cite{lany2009PRB,carey2018JPCC,elmaslmane2018JCTC,deskins2007PRB,pham2020JCTC,falletta2022npj,palermo2024PRB}. Among these, Lany and Zunger introduced a Hubbard-like functional for polarons that is nonempirically tuned to achieve the piecewise-linearity condition of the total energy \cite{lany2009PRB}. In this context, piecewise-linear DFT+$U$ functionals have been shown to lead to polaron properties in good agreement with those obtained with piecewise-linear hybrid functionals \cite{falletta2022npj}. The piecewise-linearity condition can also be enforced on each electron state through tunable orbital-dependent potentials \cite{dabo2010PRB,nguyen2018PRX,schubert2024npj}. Furthermore, two efficient semilocal functionals for polarons have been developed, namely the $\gamma$DFT functional \cite{falletta2022PRL,falletta2022PRB,falletta2023PRB} and the $\mu$DFT functional \cite{falletta2024JAP}. These functionals are based on the idea that the many-body self-interaction can be addressed by introducing a weak local potential to the semilocal Hamiltonian that depends linearly on the polaron density. Extensive comparisons between the results obtained with piecewise-linear PBE0($\alpha$), DFT+$U$, $\gamma$DFT, and $\mu$DFT functionals showed that polaron properties free from many-body self-interaction are robust with respect to the functional adopted \cite{falletta2022PRL,falletta2022PRL,falletta2022npj,falletta2023PRB,falletta2024JAP}. Considered properties include charge densities, lattice distortions, formation energies, hopping barriers, hopping rates, and hyperfine parameters. The robustness of polaron properties free from many-body self-interaction is also found with the machine-learned exchange functional CIDER \cite{bystrom2022JCTC,bystrom2024PRB,bystrom2024JCTC}, which is able to localize polarons and yield formation energies in agreement with reference hybrid functional calculations \cite{bystrom2024JCTC}. At variance, the meta-GGA functional SCAN \cite{sun2015PRL} is found to be unable to localise polarons \cite{wickramaratne2024PRB}. 

% Motivation
While developments in polaron physics from first principles are in continuous advancement  \cite{lee2021PRM,lafuentebartolome2022PRL,lafuente2022PRB,vasilchenko2022PRB,luo2022PRB,dai2024PRB}, it remains unclear how  charged density functional formulations based on the nonempirical tuning of parameters connect with neutral formulations for addressing the self-interaction \cite{sadigh2015PRB,sio2019PRB,sio2019PRL}. In particular, approaches based on nonempirical tuning such as PBE0($\alpha$), DFT+$U$, $\gamma$DFT, and $\mu$DFT, require performing calculations for a distorted supercell in the presence of an extra charge. At variance, the pSIC functional does not depend on tunable parameters and determines the polaron energetics and forces from extrapolations around the neutral charge state of the supercell with polaronic distortions \cite{sadigh2015PRB}. Alternatively, the formulation of Sio \textit{et al.}\ constructs the polaron energetics through the use of electron-phonon couplings calculated for the pristine undistorted unit cell  \cite{sio2019PRL,sio2019PRB}. However, in this unit-cell method, the energy functional is derived by neglecting electron screening effects upon electron addition, which are sizeable in the case of small polarons \cite{falletta2022PRL,falletta2022PRB}. In this context, it is important to establish a conceptual and numerical comparison between these self-interaction correction techniques. 

% content paragraph 
In this work, we demonstrate the connection and equivalence of charged and neutral density functional formulations for correcting the many-body self-interaction of polarons. We begin by focusing on charged formulations for polarons, namely PBE0($\alpha$), DFT+$U$, $\gamma$DFT, and $\mu$DFT, and illustrate the theory behind nonempirical functional tuning. Particularly, we show that the many-body self-interaction can be suppressed by achieving piecewise-linearity of the energy with respect to fractional charges. We give the expression of the formation energy calculated with a piecewise-linear functional and provide its physical interpretation. Then, we revisit the hybrid-functional formulation for the self-interaction \cite{falletta2022PRL,falletta2022PRB}, highlighting its derivation and the superiority of the many-body self-interaction over the one-body self-interaction. Next, we focus on neutral formulations for addressing the self-interaction of polarons. Specifically, by leveraging the hybrid-functional formulation for the self-interaction \cite{falletta2022PRL,falletta2022PRB}, we derive a semilocal expression for the many-body self-interaction of the polaron state. The resulting self-interaction–free functional is parameter-free and therefore does not require functional tuning. We show that this essentially coincides with the pSIC functional \cite{sadigh2015PRB}. At this point, we focus on demonstrating the connections between charged and neutral formulations for polarons. We establish a formal connection between $\gamma$DFT, $\mu$DFT, the pSIC \cite{sadigh2015PRB} functionals, as well as with the unit-cell method \cite{sio2019PRB}. Specifically, we demonstrate that these approaches lead to the same formal expression of the self-interaction corrected energy, which can be calculated using the energetics of the polaronic structure in the neutral charge state. Remaining differences between these methods are due to the applied atomic forces and the resulting polaronic distortions, which are affected by different treatments of electron screening and atomic distortions. Furthermore, we emphasize the role of finite-size corrections for defects involving ionic polarization, which noticeablky differ for charged and neutral calculations. To address concrete examples, we compare all these methods for prototypical polaron states, namely the hole polaron in MgO, the hole polaron in $\beta$-Ga$_2$O$_3$, the $V_\text{k}$ center in NaI, the electron polaron in BiVO$_4$, and the electron polaron in rutile TiO$_2$. We show that polaron properties obtained with functionals free from many-body self-interaction are in excellent agreement with each other, as found in previous related studies \cite{falletta2022PRL,falletta2022PRB,falletta2022npj,falletta2023PRB,falletta2024JAP}. We show that larger discrepancies can be found in the presence of weak polaronic bonds, such as those found for $V_\text{k}$ centers in alkali halides. Finally, we emphasize  the importance of using semilocal schemes for studying polarons as a way to efficiently perform molecular dynamics in complex systems, which may be aided by machine learning.

% organization of the work
This work is organized as follows. In Sec.\ \ref{sec:charged}, we focus on charged formulations for polarons, including the theory of nonempirical functional tuning and the hybrid-functional formulation for the self-interaction. In Sec.\ \ref{sec:neutral}, we focus on neutral formulations for polarons. Particularly, we derive a semilocal functional free from many-body self-interaction and show its equivalence with the pSIC method. We demonstrate in Sec.\ \ref{sec:equivalence} the connection between charged and neutral formulations for polarons. In Sec.\ \ref{sec:finite-size}, we discuss the importance of finite-size corrections. In Sec.\ \ref{sec:results}, we apply these methods to a variety of hole and electron polarons and compare the results. In Sec.\ \ref{sec:conclusions}, we draw the conclusions.

\section{Charged Formulation \label{sec:charged}}

\subsection{Nonempirical functional tuning \label{sec:nonempirical_tuning}}

% potential and energy expressions
We start by discussing functionals that suppress the many-body self-interaction through nonempirical tuning of their parameters, namely the hybrid functional PBE0($\alpha$) \cite{perdew1996JCP}, the Hubbard-corrected DFT+$U$ functional \cite{anisimov1991PRB,anisimov1991PRB2,anisimov1993PRB,solovyev1994PRB,czyifmmode1994PRB,liechtenstein1995PRB,anisimov1997JPCM,dudarev1998PRB,petukhov2003PRB,cococcioni2005PRB}, the $\gamma$DFT functional \cite{falletta2022PRL,falletta2022PRB,falletta2023PRB}, and the $\mu$DFT functional \cite{falletta2024JAP}. Each of these functionals relies on a parameter, which we here denote $\xi$. For the PBE0($\alpha$) functional, $\xi$ is the fraction $\alpha$ of Fock exchange admixed to the semilocal exchange. For the DFT+$U$ functional, $\xi$ is the Hubbard interaction $U$. For the $\gamma$DFT and $\mu$DFT functionals, $\xi$ denotes the strengths $\gamma$ and $\mu$ of a weak local potential added to the semilocal Hamiltonian, respectively. The Kohn-Sham equations associated with these functionals can be written as
\begin{equation} 
	\big( \mathcal  H_{\sigma}^0  + V_\sigma^{\xi}\big)\psi_{i\sigma}^{\xi} = \epsilon_{i\sigma}^{\xi} \psi_{i\sigma}^{\xi},  \label{eq:KSloc}
\end{equation}
where $\psi_{i\sigma}^\xi$ are the wave functions, $\epsilon_{i\sigma}^\xi$ are the eigenvalues, $\sigma$ is the spin, $\mathcal  H_{\sigma}^0$ is the semilocal Perdew-Burke-Ernzerhof (PBE) Hamiltonian \cite{perdew1996PRL},  and $V_\sigma^{\xi}$ is the localizing potential. For the functionals considered here, the potential $V_\sigma^\xi$ is expressed as 
\begin{align}
    \text{PBE0($\alpha$):}\quad V_\sigma^\alpha & =  -\alpha V_{\text{x}\sigma} - \alpha \sum_{i} f_{i\sigma}  \frac{\ket{\psi^\alpha_{i\sigma}}\!\bra{\psi^\alpha_{i\sigma}}}{|\mathbf{r}-\mathbf{r}'|},\\
    \text{DFT+$U$:}\quad	V_\sigma^U & = U \sum_{Imm'}  \bigg[ \frac{\delta_{mm'}}{2}    -  n_{mm'}^{I\sigma}   \bigg] \ket{\phi_{m'}^I}\!\bra{\phi_{m}^I }, \\
    \text{$\gamma$DFT:}\quad	V_\sigma^\gamma &=  q \gamma \frac{\partial V_{\text{xc}\sigma}}{\partial q}, \label{eq:vgamma} \\
    \text{$\mu$DFT:}\quad	V_\sigma^\mu &=  q \mu  n_\text{p}^\mu \delta_{\sigma,\sigma_\text{p}}, \label{eq:vmu} 
\end{align}
where $V_{\text{x}\sigma}$ is the PBE exchange potential,  $f_{i\sigma}$ is the occupation of the $i$-th orbital in the spin channel $\sigma$, $q$ is the polaron charge ($q=-1$ for electron polarons, $q=+1$ for hole polarons),  $\mathbf n^{I\sigma}$ is the occupation matrix of localized orbitals $\phi_m^I$ of state index $m$ on atom $I$, $n_\text{p}^\xi$ is the polaron density, and $\sigma_\text{p}$ is the spin channel hosting the polaron. The polaron density is obtained through $n_\text{p}^\xi = |\psi_\text{p}^\xi|^2$, where $\psi_\text{p}^\xi$ is the wave function of the polaron state, namely the highest-occupied energy state for electron polarons, or the lowest-unoccupied energy state for hole polarons. The total energy corresponding to Eq.\ \eqref{eq:KSloc} is given by
\begin{equation}
	E^\xi = E^0  + \Delta E^\xi,
\end{equation}
where $E^0$ is the semilocal PBE energy, and  $\Delta E^\xi$ is the energy contribution corresponding to the potential $V_\sigma^\xi$. Specifically,
\begin{align}
    \text{PBE0($\alpha$):}\quad	\Delta E^\alpha & =  -\alpha E_\text{x}[n_{\uparrow}^\alpha,n_\downarrow^\alpha]  + \alpha E_\text{X}[\{\psi^\alpha_{i\sigma}\}],  \\
    \text{DFT+$U$:}\quad \Delta  E^U & =  \frac{U}{2} \sum_{I\sigma}  \text{Tr} [  \textbf{n}^{I\sigma}(1 - \textbf{n}^{I\sigma})], \\
    \text{$\gamma$DFT:}\quad	\Delta  E^\gamma &= \frac{q}{2}\sum_{\sigma} \!\int\!\!d\mathbf r  \, V^\gamma_\sigma(\mathbf r) \frac{dn^\gamma_\sigma(\mathbf r)}{dq}, \label{eq:egamma} \\
    \text{$\mu$DFT:}\quad \Delta  E^\mu &=\frac{q}{2} \sum_{\sigma}  \!\int\!\!d\mathbf r  \, V^\mu_\sigma(\mathbf r)  \frac{dn_{\sigma}^\mu(\mathbf r)}{dq}, \label{eq:emu} 
\end{align}
where $E_\text{x}$ is the semilocal exchange energy, $E_\text{X}$ is the Fock exchange energy, $n_\sigma^\xi$ is the total electron density in the spin channel $\sigma$. The $\gamma$DFT and $\mu$DFT expressions for the energy and Hamiltonian corrections are variationally related when assuming that the total energy depends quadratically on the polaron charge $q$ \cite{falletta2023PRB,falletta2024JAP}. In addition, we note that, in the  $\gamma$DFT and $\mu$DFT schemes, the localizing potentials and the corresponding energy terms vanish in the absence of polaron charge. This implies that, for $q=0$, the $\gamma$DFT and $\mu$DFT functionals reduce to the standard PBE functional \cite{falletta2022PRL,falletta2022PRB,falletta2023PRB,falletta2024JAP}. Hence, $\gamma$DFT and $\mu$DFT yield the same  valence and conduction band levels as the PBE functional. To avoid possible resonances between the polaron state and the band edge states, one can further add a self-consistent scissor operator to the Hamiltonian to shift the band states by an arbitrary energy \cite{falletta2023PRB}. 

% tuning of the parameter to enforce the PWL
The parameter $\xi$ directly influences the self-interaction of the polaron state. At $\xi=0$, all the functionals mentioned above reduce to the PBE functional, which typically results in charge delocalization due to the positive concavity of the total energy as a function of the fractional number of electrons. As $\xi$ increases, polaron localization is increasingly favored. For instance, in the case of the PBE0($\alpha=1$) functional, which essentially coincides with the Hartree-Fock functional, the many-body self-interaction is overcorrected and the  total energy exhibits a negative concavity with respect to  fractional number of electrons. Hence, one can find a value  $\xi = \xi_\text{k}$ such that the second derivative of the total energy with respect to $q$ vanishes, whereby  the piecewise-linearity condition is enforced. Through Janak's theorem \cite{janak1978PRB}, this implies that the polaron energy level $\epsilon_\text{p}^\xi$ remains constant when changing the occupation of the polaron state, specifically 
\begin{equation}
	\left.\frac{d^2}{dq^2} E^{\xi}(q)\right|_{\xi = \xi_\text{k}} \stackrel{\text{Janak}}{=} - \left.\frac{d}{dq} \epsilon_\text{p}^{\xi}(q)\right|_{\xi = \xi_\text{k}}= 0, \label{eq:PWL}
\end{equation}
where the derivatives are taken at a fixed structure with  polaronic lattice distortions. Assuming that the total energy is quadratic with respect to the polaron charge \cite{falletta2022PRB}, the condition in Eq.\ \eqref{eq:PWL} is met when the neutral and charged polaron levels are identical, i.e., $\epsilon_\text{p}^{\xi_\text{k}}(q) = \epsilon_\text{p}^{\xi_\text{k}}(0)$, where $q=\pm 1$.  In Eq.\ \eqref{eq:PWL}, the electronic structure and the distorted polaronic geometry are self-consistently calculated with the functional with parameter $\xi=\xi_\text{k}$. We emphasize that self-trapped electron polarons involve the addition of an electron ($q = -1$), while self-trapped hole polarons involve the removal of an electron ($q=+1$) from the system. Typically, in electronic structure codes, electrons are added in the spin up channel, and removed from the spin down channel. Hence, with this convention, in the case of electron polarons at finite $q$, the energy level $\epsilon^\xi_\text{p}(q)$ corresponds to the highest occupied energy level below the conduction band in the spin up channel. At $q=0$, $\epsilon_\text{p}^\xi(0)$ is the lowest unoccupied energy level above the valence band in the spin up channel.  Analogously, in the case of hole polarons, $\epsilon_\text{p}^\xi(+1)$ corresponds to the lowest unoccupied energy level above the valence band in the spin down channel. For fractional $q$ and for $q=0$, $\epsilon_\text{p}^\xi(q)$ is the highest occupied energy level above the valence band in the spin down channel. Here, the states at $q=\pm 1$ and at $q=0$ are calculated using a supercell with polaronic lattice distortions. Hence, the state at $q=\pm 1$  corresponds to the charged polaronic system, while the state at $q=0$ corresponds to the neutral system with distorted polaron structure.

% practical way the parameter is found
The practical procedure for localizing polarons and determining $\xi_\text{k}$ is outlined in the following. First, one needs to initialize polaronic-like lattice distortions in the supercell under consideration. This is crucial as structural relaxations with an extra charge in a pristine symmetric system would unlikely lead to structural symmetry breaking. There are various ways by which one can construct the initial lattice distortions, e.g.,  by distorting lattice bonds according to chemical intuition or by performing relaxations with substitutional atoms that mimic the presence of an extra charge at a given site.  Once the initial structure established, one needs to provide an initial guess for the parameter $\xi$ that can lead to polaron localization. Generally, one performs structural relaxations for increasing values of $\xi$ until a structural relaxation converging towards a distorted system is found. Polaron localization can be verified by inspecting the density of the polaron orbital or the spin density of the system. Once the polaron localization is obtained for a given value of $\xi$,  higher values of $\xi$ generally still yield localized polarons. In this way, one can identify a set of $\xi$ values, calculate the respective polaron structures, and determine the polaron energy levels at charge $q=-1$ for electron polarons or $q=+1$ for hole polarons, and at charge $q=0$ for the neutral distorted system with the polaron structure. Then, through Eq.\ \eqref{eq:PWL}, one can verify the extent to which the piecewise-linearity condition is enforced. As the dependence of the energy levels on $\xi$ is approximately  linear  \cite{falletta2022PRB}, it is straightforward to find the value $\xi_\text{k}$ such that the charged and neutral polaron levels coincide, as required for suppressing the many-body self-interaction.

% polaron formation energy
The polaron stability can be quantified through the concept of formation energy, which measures the energy gain due to polaron localization compared to the case of a delocalized charge in the undistorted system. This is obtained by taking the difference between the total energy $E^\xi(q)$ of the polaron system, and the total energy of a pristine undistorted system with an extra charge. The latter can be calculated as $E_\text{ref}^\xi(0) - q \epsilon_\text{b}^\xi$, where $E_\text{ref}^\xi(0)$ is the total energy of the neutral undistorted system, and  $\epsilon_\text{b}^\xi$ is the energy level of the delocalized band-edge state. The delocalized band-edge state refers to the bottom of  the  conduction band for electron polarons, and to the top of the valence band for hole polarons. This leads to the following expression for the formation energy \cite{freysoldt2014RMP}:
\begin{align}
	E_\text{f}^{\xi}(q)  = E^\xi(q) - E_\text{ref}^\xi(0) + q \epsilon_\text{b}^\xi. \label{eq:Ef}
\end{align}
When the functional is tuned in order to enforce the piecewise-linearity condition in Eq.\  \eqref{eq:PWL},  Eq.\ \eqref{eq:Ef} can be conveniently rewritten as \cite{falletta2022PRB,falletta2022PRL,yuan2019PRB}
\begin{align}
	E_\text{f}^{\xi_\text{k}}(q)  = q(\epsilon_\text{b}^{\xi_\text{k}} - \epsilon_\text{p}^{\xi_\text{k}} ) + [E^{\xi_\text{k}}(0) - E_\text{ref}^{\xi_\text{k}}(0)]. \label{eq:Ef_2}
\end{align}
The expression in Eq.\ \eqref{eq:Ef_2} carries a transparent physical interpretation. The first term consisting of the difference between band and polaron energy levels represents the energy gain due to charge localization. The second term written as difference between the neutral distorted and the neutral undistorted systems quantifies the energy cost due to lattice distortions. Equation \eqref{eq:Ef_2} establishes a key connection among all piecewise-linear functionals [PBE0($\alpha$), DFT+$U$, $\gamma$DFT, and $\mu$DFT], leading to the robustness of polaron formation energies obtained with functionals free from many-body self-interaction \cite{falletta2022PRB,falletta2022PRL}.

\vspace*{1cm}
\subsection{Unified hybrid-functional formulation for the self-interaction  \label{sec:unified_hybrid}}

% intro to the unified hybrid-functional formulation
In this section, we summarize the main results of the unified hybrid-functional formulation for the self-interaction introduced in Refs.\ \cite{falletta2022PRL,falletta2022PRB}. This formulation is based on the PBE0($\alpha$) hybrid functional, which encompasses the  cases of the semilocal PBE functional ($\alpha=0$), the Hartree-Fock  functional ($\alpha=1$), and the piecewise-linear functional ($\alpha = \alpha_\text{k}$). This is instrumental in order to establish a connection between one-body ($\alpha=1$) and many-body ($\alpha=\alpha_\text{k}$) forms of self-interaction and to formally demonstrate the superiority of the many-body notion of self-interaction over the one-body notion of self-interaction. The formulation is based on the assumption that the total energy is linear in $\alpha$ and quadratic in the polaron charge $q$. This implies neglecting variations of the wave functions $\psi_{i\sigma}^\alpha$ with $\alpha$, variations of the polaron density $n^\alpha_\text{p}$ with $q$, and second-order variations of the density of valence electrons $n^\alpha_{\sigma\text{val}}$ with $q$, namely
\begin{align}
\frac{d\psi^\alpha_{i\sigma}}{d\alpha} = 0, \quad 	\frac{dn^\alpha_\text{p}}{dq}  = 0, \quad \frac{d^2n^\alpha_{\sigma\text{val}}}{dq^2} = 0, \label{eq:assumptions}
\end{align}
as demonstrated in Appendix A of Ref.\ \cite{falletta2022PRB}. Hence, in the following, we can omit the superscript $\alpha$ from the notation of the wave functions for simplicity.

% derivation and many-body self-interaction 
The many-body self-interaction is defined as deviation of the total energy from the piecewise-linear dependence upon addition of fractional  charge. The PBE0($\alpha$) becomes piecewise linear in correspondence of a fraction $\alpha=\alpha_\text{k}$ of Fock exchange. Hence, through Janak's theorem \cite{janak1978PRB}, one can take the polaron level $\epsilon_\text{p}^{\alpha_\text{k}}$ as reference slope of the energy with respect to the fractional number of electrons. Then, the many-body self-interaction energy correction can be written as
\begin{equation}
    \left.\Delta E^{\alpha}(q)\right|_\text{mb}^\text{hyb} = [E^\alpha(0) - q \epsilon_\text{p}^{\alpha_\text{k}}] - E^\alpha(q),  \label{eq:esic} \vspace*{0.1cm}
\end{equation}
which measures the difference between the straight-line and the quadratic dependence of the total energy obtained with a given PBE0($\alpha$) functional. Using the assumptions of this formulation [cf.\ Eq.\ \eqref{eq:assumptions}], we can rewrite the latter expression as follows \cite{falletta2022PRB}:
\begin{widetext}
\begin{equation}
    \left.\Delta E^\alpha(q)\right|_\text{mb}^\text{hyb} = -  \Big(  1-\frac{\alpha}{\alpha_\text{k}} \Big)    \Big[  (q-q_\text{k})^2  - q_\text{k}^2   \Big]   \Bigg\lbrace E_\text{H}\bigg[\frac{dn}{dq}\bigg]  + \frac{1}{2} \sum_{\sigma \sigma'}\!\int\!\! d\mathbf r d\mathbf r'\, \frac{\delta^2 E_\text{xc}[n_\uparrow,n_\downarrow]}{\delta n_{\sigma}(\mathbf r) \delta n_{\sigma'}(\mathbf r')} \frac{dn_\sigma(\mathbf r)}{dq} \frac{dn_{\sigma'}(\mathbf r)}{dq}  \Bigg\rbrace, \label{eq:mbSIE}
\end{equation}
\end{widetext}
where $q_\text{k}$ is a fractional charge that such that $\epsilon_\text{p}^\alpha(q_\text{k}) = \epsilon_\text{p}^{\alpha_\text{k}}$ for every $\alpha$, and is defined as follows:
\begin{equation}
q_\text{k} = - \frac{d\epsilon_\text{p}^\alpha(0)}{d\alpha} \Bigg/ \frac{d^2 \epsilon_\text{p}^\alpha(q)}{d\alpha dq}. \label{eq:qk_alpha}
\end{equation}
The roles of $\alpha_\text{k}$ and $q_\text{k}$ are dual in determining the energy level $\epsilon_\text{p}^{\alpha_\text{k}}$ \cite{falletta2022PRL,falletta2022PRB}, which is obtained with the piecewise-linear PBE0($\alpha_\text{k}$) hybrid functional. Specifically, the polaron level $\epsilon_\text{p}^{\alpha_\text{k}}$ is obtained for every charge $q$ at $\alpha=\alpha_\text{k}$ and for every $\alpha$ at charge $q=q_\text{k}$. The functional expression in Eq.\ \eqref{eq:mbSIE} is semilocal, as it only depends on Hartree and exchange-correlation energy terms. The complexity of the Fock operator is encoded in the parameters $\alpha_\text{k}$  and $q_\text{k}$. In the case of a PBE calculation ($\alpha=0$), the dependence on $\alpha_\text{k}$ vanishes and the expression can be further simplified \cite{falletta2022PRL}. We remark that the derivatives $dn_\sigma/dq$ in Eq.\ \eqref{eq:mbSIE} represent the variations of the total electron density in the spin channel $\sigma$, which are found to be sizeable and nonnegligible for small polarons of both electron and hole kinds \cite{falletta2022PRB,falletta2022PRL}.

% one-body self-interaction and superiority of the many-body self-interaction
The one-body self-interaction energy correction can be expressed as the difference between the concavity obtained for the Hartree-Fock like functional at $\alpha=1$ and the concavity obtained for a given value of $\alpha$, namely
\begin{equation}
	\left.\Delta E^{\alpha}(q)\right|_\text{ob}^\text{hyb} = \big[E^{1}(q) - E^{1}(0)\big] - \big[E^{\alpha}(q) - E^{\alpha}(0)\big]. \label{eq:deriv_ob}
\end{equation}
It can be shown that the above expression can be related to that of the many-body self-interaction correction by a simple relation \cite{falletta2022PRB}, namely 
\begin{equation}
    \left.\Delta E^\alpha(q)\right|_\text{mb}^\text{hyb} =
    \frac{\alpha_\text{k} - \alpha}{1 - \alpha} \left.\Delta E^\alpha(q)\right|_\text{ob}^\text{hyb}, \label{eq:link_SI_alfa}
\end{equation}
which quantifies the relation between many-body and one-body forms of self-interaction in terms of $\alpha$ and $\alpha_\text{k}$. In the case of a PBE functional ($\alpha=0$), Eq.\ \eqref{eq:link_SI_alfa} can be rewritten as
\begin{equation}
    \left.\Delta E^{0}(q)\right|_\text{mb}^\text{hyb} \simeq \frac{1}{\varepsilon_\infty}  \left.\Delta E^{0}(q)\right|_\text{ob}^\text{hyb},
    \label{eq:rel_SI_pbe2}
\end{equation}
where we approximated $\alpha_\text{k}\simeq1/\varepsilon_\infty$ to reproduce the correct asymptotic potential in the long-range limit \cite{refaely2015PRB,kronik2020PCCP}. In addition,  $\alpha=1/\varepsilon_\infty$ typically yields band gaps in good agreement with experiment 
\cite{skone2014PRB,miceli2018PRB,chen2018PRM,bischoff2019PRM} upon consideration of all relevant effects, such as spin-orbit coupling, renormalization due to vibrations, and excitonic and magnetic ordering effects \cite{wiktor2017PRM,wiktor2018JPCL,capano2020JPCC,falletta2020JMCA}. Equation \eqref{eq:rel_SI_pbe2} highlights physical intuition behind the two notions of self-interaction. In particular, it shows that while both one-body and many-body forms of self-interaction account for screening effects in the electron density, the many-body self-interaction also includes screening effects in the Coulomb kernel, which  are not accounted for by the one-body self-interaction. This carries similarities with the Hartree-Fock theory of excitons \cite{bassani1976book,henderson2011PSS}, 
where the exciton binding energy depends on the screened Coulomb kernel $1/(\epsilon_\infty|\mathbf r - \mathbf r'|)$ and not on the bare Coulomb kernel  $1/|\mathbf r - \mathbf r'|$. This demonstrates the superiority of the many-body self-interaction over the one-body self-interaction.

\section{Neutral Formulation\label{sec:neutral}}

\subsection{Analytical expression of the many-body self-interaction in semilocal DFT \label{sec:semilocal}}

% limitations of the hybrid-functional formulation
The unified hybrid-functional formulation for the self-interaction highlights the importance of correcting the many-body self-interaction with respect to the one-body self-interaction. However, this formulation carries practical limitations for efficient calculations of polarons at the semilocal level of theory. Indeed, despite the semilocal nature of the expression in Eq.\ \eqref{eq:mbSIE}, hybrid-functional calculations cannot be fully avoided when obtaining polaron formation energies. For this, one needs to determine the energy separation between the polaron level and the delocalized band edge, which gives the energy gain due to charge localization [cf.\  Eq.\ \eqref{eq:Ef_2}]. This would require not only the knowledge of the polaron level at $\alpha=\alpha_\text{k}$ but also the corresponding delocalized band edge at $\alpha=\alpha_\text{k}$. In addition, the expression in Eq.\ \eqref{eq:mbSIE} depends on the parameters $\alpha_\text{k}$ and $q_\text{k}$, which intrinsically depend on the application of  the Fock operator \cite{falletta2022PRB}. While $\alpha_\text{k}$ can efficiently be determined through the use of hydrogen probes \cite{bischoff2019PRB,bischoff2019PRM,bischoff2021PRR}, it remains unclear how to determine the value of $q_\text{k}$ without resorting to hybrid functional calculations on the polaron supercell.  Hence, practically, the unified formulation described in  Sec.\ \ref{sec:unified_hybrid} cannot straightforwardly be employed for efficient studies of polarons at the semilocal level of theory \cite{falletta2022PRL,falletta2022PRB}. 

% intuition from gammaDFT and muDFT
In consideration of  these limitations, the $\gamma$DFT  and $\mu$DFT schemes have been introduced to enable efficient calculations of polarons at the semilocal level of theory. These functionals are based on the addition of a weak local potential to the semilocal PBE Hamiltonian to suppress the many-body self-interaction [cf.\ Eqs.\ \eqref{eq:vgamma} and \eqref{eq:vmu}]. Interestingly, when $q=0$, the extra localizing potential vanishes and the Hamiltonian falls back to the PBE Hamiltonian. This implies three important results. First, the $\gamma$DFT and $\mu$DFT band edges coincide with those obtained with PBE, namely $\epsilon_\text{b}^\xi = \epsilon_\text{b}^0$. Second, the $\gamma$DFT and $\mu$DFT polaron levels obtained in the absence of charge ($q=0$) but in the presence of polaronic distortions coincide with their respective PBE value, i.e., $\epsilon_\text{p}^\xi(0) = \epsilon_\text{p}^0(0)$ for any $\xi$. Third, the many-body self-interaction corrected polaron level coincides with the neutral polaron level calculated with the PBE functional. This arises from the fact that $\xi = \xi_\text{k}$ enforces the piecewise-linearity condition in Eq.\ \eqref{eq:PWL},  and hence the charged and the neutral polaron level coincide, namely $\epsilon_\text{p}^{\xi_\text{k}} = \epsilon_\text{p}^{\xi_\text{k}}(0) = \epsilon_\text{p}^{0}(0)$.  Through Janak's theorem \cite{janak1978PRB}, the total energy can be written as:
\begin{equation}
	E^{\xi_\text{k}}(q) = E^{\xi_\text{k}}(0) -  \int_0^q dq' \epsilon_\text{p}^{\xi_\text{k}}(q').
\end{equation}
On the basis of the above arguments, the total energy $E^{\xi_\text{k}}(0)$ of the neutral system with polaronic distortions coincides with the PBE energy, i.e., $E^{0}(0)$. Then, using the fact that $\epsilon_\text{p}^{\xi_\text{k}}(q') = \epsilon_\text{p}^0(0)$,  the total energy of the polaron system at $\xi = \xi_\text{k}$ can be rewritten as
\begin{equation}
    E^{\xi_\text{k}}(q) = E^{0}(0) - q \epsilon_\text{p}^0(0), \label{eq:E_PWL_semilocal}
\end{equation}
where all  quantities on the right hand side are calculated at the PBE level of theory. This implies that, once the polaron structure is provided, the polaron energetics can be calculated with only PBE energies and energy levels.

% derivation of the semilocal formulation
Following the same assumptions and logic behind the hybrid-functional formulation in Sec.\ \ref{sec:unified_hybrid} on the quadratic dependence of the energy functional $E^0(q)$ on $q$, namely
\begin{align}
	\frac{dn_\text{p}}{dq}  = 0, \quad \frac{d^2n_{\sigma\text{val}}}{dq^2} = 0, \label{eq:assumptions_pbe}
\end{align}
and inspired by Eq.\ \eqref{eq:E_PWL_semilocal}, we define the many-body self-interaction correction as
\begin{equation}
    \left.\Delta E^{0}(q)\right|_\text{mb}^\text{sl} = [E^0(0) - q \epsilon_\text{p}^0(0)] - E^0(q),  \label{eq:esic_sl} \vspace*{0.1cm}
\end{equation}
which measures the deviation from the piecewise-linear behavior of the PBE total energy.  Then, using Eq.\ \eqref{eq:assumptions_pbe}, we expand the Eq.\ \eqref{eq:esic_sl} to second order in $q$:
\begin{widetext}
\begin{equation}
    \left.\Delta E^{0}(q)\right|_\text{mb}^\text{sl} = - \frac{q^2}{2} \frac{d^2 E^0(q)}{dq^2}  = - q^2   \Bigg\lbrace E_\text{H}\bigg[\frac{dn}{dq}\bigg]  + \frac{1}{2} \sum_{\sigma \sigma'}\!\int\!\! d\mathbf r d\mathbf r'\, \frac{\delta^2 E_\text{xc}[n_\uparrow,n_\downarrow]}{\delta n_{\sigma}(\mathbf r) \delta n_{\sigma'}(\mathbf r')} \frac{dn_\sigma(\mathbf r)}{dq} \frac{dn_{\sigma'}(\mathbf r)}{dq}  \Bigg\rbrace, \label{eq:mbSIE_sl}
\end{equation}
\end{widetext}
which represents the many-body self-interaction energy correction at the semilocal PBE level of theory. In Eq.\ \eqref{eq:mbSIE_sl}, electron screening effects are accounted for through the derivatives of the total densities with respect to $q$, namely $dn_\sigma/dq$. In addition, the expression in Eq.\ \eqref{eq:mbSIE_sl} is parameter-free. 

In Eq.~\eqref{eq:mbSIE_sl}, the derivatives $dn_\sigma/dq$ account for the presence of the polaron state and for the first-order variations in the valence electron density upon addition of the polaron, thereby capturing first-order screening effects. Second-order screening effects, represented by the second derivatives of the valence electron density with respect to $q$, as well as variations of the polaron density with $q$, are neglected [cf.~Eq.~\eqref{eq:assumptions_pbe}]. We note that when using the PBE functional, the polaron often delocalizes as $q$ varies from $0$ to $\pm1$, thus deviating from the assumption that the polaron density is independent of the polaron charge. However, when the energy term in Eq.~\eqref{eq:mbSIE_sl} is subtracted from the PBE energy, the polaron remains localized, and variations of its density can be neglected, similarly to the case of PBE0($\alpha$) functionals [cf.\ Eq.\ \eqref{eq:assumptions}]. Therefore, the energy correction in Eq.~\eqref{eq:mbSIE_sl} remains valid when properly removed from the PBE total energy in the minimization of the Kohn–Sham equations.

% corresponding potential correction and forces
Given the expression of self-interaction correction energy in Eq.\ \eqref{eq:mbSIE_sl}, we can determine the corresponding self-interaction potential  correction to be added to the PBE Hamiltonian for suppressing the many-body self-interaction. Applying to Eq.\ \eqref{eq:mbSIE_sl} the variational relation between the Kohn-Sham potential and the Hamiltonian, namely $V^0_\sigma(\mathbf r) = \delta E^0/\delta n_\sigma(\mathbf r)$,  leads to the following expression for the self-interaction potential correction:
\begin{widetext}
	\begin{equation}
		\left.\Delta V_\sigma^{0}(q)\right|_\text{mb}^\text{sl}  = - q   \Bigg\lbrace V_\text{H}\bigg[\frac{dn}{dq}\bigg]  +  \sum_{\sigma'}\!\int\!\! d\mathbf r'\, \frac{\delta V_{\text{xc}\sigma}(\mathbf r)}{\delta n_{\sigma'}(\mathbf r')} \frac{dn_{\sigma'}(\mathbf r)}{dq}  \Bigg\rbrace = - q \frac{d\mathcal H^0_\sigma(q)}{dq} ,\label{eq:mbSIC_pot_sl}
	\end{equation}
\end{widetext}
where we expressed the self-interaction correction to the semilocal Hamiltonian as the first derivative of the Hamiltonian $\mathcal H_\sigma^0$ with respect to $q$.

% comparison with the hybrid-functional formulation
There are evident similarities between the expression in Eq.\ \eqref{eq:mbSIE_sl} and that in Eq.\ \eqref{eq:mbSIE} obtained in the hybrid-functional formulation for the self-interaction. In particular, the result in Eq.\ \eqref{eq:mbSIE_sl} coincides with that in Eq.\ \eqref{eq:mbSIE} when $\alpha = 0$ and $q_\text{k} = 0$.
Here, $\alpha = 0$ is due to the fact that $\left.\Delta E^{0}(q)\right|_\text{mb}^\text{sl}$ corrects the PBE total energy, to which the general PBE0($\alpha$) formulation falls back when $\alpha=0$. On the other hand, $q_\text{k} = 0$ can be explained as follows. Following Eq.\ \eqref{eq:qk_alpha}, we can generally define  $q_\text{k}$ as a function of $\xi$ via \cite{falletta2022PRB}
\begin{equation}
	q_\text{k} = - \frac{d\epsilon_\text{p}^\xi(0)}{d\xi} \Bigg/ \frac{d^2 \epsilon_\text{p}^\xi(q)}{d\xi dq}, 
 \label{eq:qk_xi}
\end{equation}
where $\epsilon_\text{p}^\xi$ is the polaron level obtained with a tunable functional parametrized by $\xi$, as discussed in Sec.\ \ref{sec:nonempirical_tuning}. 
Hence, $q_\text{k}=0$ in this formulation signifies that the neutral polaron level $\epsilon_\text{p}^\xi(0)$ in the numerator of Eq.\ \eqref{eq:qk_xi} does not 
depend on $\xi$. This corresponds to the description achieved in $\gamma$DFT and $\mu$DFT, where the neutral polaron level and the corresponding band-edge level are independent of $\xi$ by construction, as the extra potential vanishes at $q=0$ [cf.\ Eqs.\ \eqref{eq:vgamma} and \eqref{eq:vmu}].
Similarly, in PBE0($\alpha$) calculations, the energy separation between the neutral polaron level  $\epsilon_\text{p}^\alpha(0)$ and the corresponding 
band-edge level barely depends on $\alpha$ \cite{falletta2022PRL,falletta2022PRB}. However, in this case the overall shift of the band edge as a function of $\alpha$ results in a finite value of $q_\text{k}$. Instead, the semilocal expression in Eq.\ \eqref{eq:mbSIE_sl} does not depend on any tunable parameter or on any parameter involving the Fock operator, thereby overcoming the issues of functional tuning and the limitations of the formulation based on hybrid functionals (Sec.\ \ref{sec:unified_hybrid}).   

\subsection{Neutral functional for correcting the self-interaction in semilocal DFT \label{sec:sicDFT}}

We show that the expression of the many-body self-interaction corrections derived in Sec.\ \ref{sec:semilocal} can be used to construct a self-interaction corrected energy functional, which we then connect to the pSIC functional. Specifically, Eqs.\ \eqref{eq:mbSIE_sl} and \eqref{eq:mbSIC_pot_sl} define the self-interaction corrections for the PBE total energy and Hamiltonian, respectively. By including the correction term in Eq.\ \eqref{eq:mbSIE_sl} to the PBE energy functional, one can define the self-interaction corrected energy functional as
\begin{align}
    E^\text{sic}(q) & = E^{0}(0) + q \left.\frac{dE^0(q)}{dq}\right|_0 +  \frac{q^2}{2} \frac{d^2 E^0(q)}{dq^2} + \left.\Delta E^{0}(q)\right|_\text{mb}^\text{sl}, \nonumber \\ 
    & = E^{0}(0) + q \left.\frac{dE^0(q)}{dq}\right|_0, \label{eq:esic_0_first}
\end{align}
which coincides with the Taylor expansion of the energy with respect to the polaron charge $q$ up to first order, the second order derivative having been removed since  it represents the many-body self-interaction energy [cf.\ Eq.\ \eqref{eq:mbSIE_sl}]. We note that the first derivative of the energy with respect to the polaron charge is taken at fixed structure with polaronic lattice distortions, and thus coincides with the energy level of the neutral polaron state calculated at the PBE level of theory, namely $\left. dE^0(q)/ dq \right|_0 = - \epsilon_\text{p}^0(0)$, as follows from Janak's theorem \cite{janak1978PRB}.

By adding  the potential correction in Eq.\ \eqref{eq:mbSIC_pot_sl} to the PBE Hamiltonian, we find the first-order dependence on the charge $q$ to be exactly cancelled, resulting in the following expression for the self-interaction corrected Hamiltonian:
\begin{align}
	\mathcal H_\sigma^\text{sic}(q) & = \mathcal H_\sigma^0(0) + q \frac{d\mathcal H^0_\sigma(q)}{dq} + \left.\Delta V_\sigma^{0}(q)\right|_\text{mb}^\text{sl} = \mathcal H_\sigma^0(0), \label{eq:Hsic_0_first}
\end{align}
where higher order terms in $q$ vanish because the total energy is assumed to be quadratic in $q$.
Hence, $\mathcal H_\sigma^\text{sic}(q)$ is found to coincide with the PBE Hamiltonian of the neutral system with polaronic distortions, namely $\mathcal H^0_\sigma(0)$. 

Through the Hellmann–Feynman theorem, the atomic forces corresponding to the energy functional in Eq.\ \eqref{eq:esic_0_first} can be written as
\begin{equation}
	F_{i\nu}^\text{sic}(q) = F_{i\nu}^{0}(0) + q \frac{dF_{i\nu}^0(q)}{dq}, \label{eq:forces_sic}
\end{equation}
where $F_{i\nu}^{0}(0)$ is the Cartesian component $\nu$ of the PBE force acting on atom $i$ in the absence of charge ($q=0$), and the derivative is taken at fixed structure with polaronic lattice distortions. The forces $F_{i\nu}^{0}(0)$  can be easily calculated through analytical expressions by carrying out a calculation at $q=0$ for the system with polaronic distortions. The first-order derivative of the forces in Eq.\ \eqref{eq:E_PWL_semilocal} can be obtained by finite differences. The simplest forward finite-difference expression
gives:
\begin{equation}
	\frac{dF_{i\nu}^0(q)}{dq} = \frac{F_{i\nu}^0(\delta q) - F_{i\nu}^0(0)}{\delta q}, \label{eq:fSIC_ff}
\end{equation}
where $\delta q$ is an infinitesimal charge, and $F_{i\nu}^0(\delta q)$ and $F_{i\nu}^0(0)$ are determined for the same fixed polaron structure. The infinitesimal charge $\delta q$ is chosen to be sufficiently large to avoid numerical problems and sufficiently small to guarantee the accuracy of the finite-difference expression. In particular, $\delta q$ is negative for electron polarons and positive for hole polarons. Practically, higher-order finite-difference expressions can be used to converge the derivative of the forces with respect to $q$. We remark that the forces defined in Eq.\ \eqref{eq:forces_sic} depend on derivatives with respect to $q$.  Through these derivatives, the screening effects in response to infinitesimal charges can be captured. Indeed, in the absence of such derivatives, the forces would solely depend on the PBE Hamiltonian at $q=0$ and the screening effects of  would be missed.

% formation energy
The energy and forces in Eqs.\ \eqref{eq:esic_0_first} and \eqref{eq:forces_sic} enable electronic and structural relaxations of polarons without the need of functional tuning. Then, the polaron formation energy free from many-body self-interaction can be calculated as
\begin{align}
    E_\text{f}^\text{sic}(q)  = q[\epsilon_\text{b}^{0} - \epsilon_\text{p}^{0}(0) ] + [E^{0}(0) - E_\text{ref}^{0}(0)], \label{eq:Ef_2_sl}
\end{align}
which is defined in terms of total energies and energy levels solely calculated at the PBE level of theory. Specifically, the energy gain due to charge localization is measured as difference between the PBE energy level of the neutral polaron state and the PBE energy level of the delocalized band-edge state. Similarly, the cost due to lattice distortions is calculated as the difference between the PBE total energy of the neutral system with polaronic distortions and the PBE total energy of the neutral pristine system. The fact that the band gap is generally underestimated in PBE does not pose any limitation in the prediction of the ground state or of the transport properties of polarons, as the energy gain due to charge localization depends only on the relevant band manifold  and only weakly involves the  band states on the opposite side of the band gap [cf.\ Eq.\ \eqref{eq:Ef_2_sl}]. This is further supported by the robustness of the polaron formation energies obtained when using piecewise-linear functionals \cite{falletta2022PRL,falletta2022PRB,falletta2022npj,falletta2023PRB,falletta2024JAP}.

\subsection{The pSIC functional}

% energetics
We now compare the functional derived in Sec.\ \ref{sec:sicDFT} with the  pSIC functional introduced by Sadigh, Erhart, and \AA{}berg \cite{sadigh2015PRB}. In the pSIC scheme, the energy functional is written as
\begin{equation}
	E^\text{pSIC}(q) = E^{0}(0) + q \left.\frac{dE^0(q)}{dq}\right|_0 + \Pi_\text{p},
	\label{eq:e_psic}
\end{equation}
where $\Pi_\text{p}$ is a correction to the total energy coming from contributions beyond PBE. The correction  $\Pi_\text{p}$  leads to a shift $\Delta_\text{p}$ of the neutral polaron level, which  is calculated through $G_0W_0$ in Ref.\ \cite{sadigh2015PRB}. We remark that the correction  $\Delta_\text{p}$ of the polaron level should be accompanied with an analogous correction
$\Delta_\text{b}$ to the corresponding delocalized band edge. Hence, the pSIC energy levels of the polaron state and of the delocalized band edge are given by
\begin{align}
	\epsilon_\text{p}^\text{pSIC} (0)&= \epsilon_\text{p}^0(0) + \Delta_\text{p} , \\
	\epsilon_\text{b}^\text{pSIC} &= \epsilon_\text{b}^0 + \Delta_\text{b} .
\end{align}
In Ref.\ \cite{sadigh2015PRB}, $\Delta_\text{p}$ and $\Delta_\text{b}$ are assumed to be independent of the structure and are taken to coincide. This leads to the following polaron formation energy:
\begin{equation}
	E_\text{f}^\text{pSIC}(q)  = q[\epsilon_\text{b}^{0} - \epsilon_\text{p}^{0}(0) ] + [E^{0}(0) - E_\text{ref}^{0}(0)], \label{eq:Ef_2_sl_psic}
\end{equation}
where  $\epsilon_\text{b}^\text{pSIC}(0) - \epsilon_\text{p}^\text{pSIC}(0) = \epsilon_\text{b}^0 - \epsilon_\text{p}^0$ under the assumption  that $\Delta_\text{p} = \Delta_\text{b}$. In this case, $\Pi_\text{p}$ is approximated as independent of the structure and thus does not lead to any additional force to the usual Hellmann-Feynmann forces. The forces are then determined through finite-differences, as in Eqs.\ \eqref{eq:forces_sic} and \eqref{eq:fSIC_ff}.

% difference sicDFT and pSIC
The main difference between the functional derived in Sec.\ \ref{sec:sicDFT} and pSIC consists in the inclusion of the $G_0W_0$ corrections to the energetics in pSIC. However, in pSIC, these corrections do not affect the forces and the polaron formation energy. Hence, practically, the functional derived in Sec.\ \ref{sec:sicDFT} and pSIC coincide. In the following, we thus conform to the terminology introduced previously in the literature and denote this functional pSIC.

\subsection{The unit-cell method}

We now discuss the unit-cell method for polarons introduced by Sio \textit{et al.}\ \cite{sio2019PRB,sio2019PRL}, which is based on the use of a neutral undistorted unit cell. The derivation of this method, along with its underlying assumptions, was primarily developed for large polarons, which are more challenging to simulate because of the high computational cost associated with large supercell calculations. We briefly outline the derivation of the unit-cell energy functional, using the notation adopted in the present work. In Ref.\ \cite{sio2019PRB}, the energy functional is assumed to depend quadratically on the polaron density. This is analogous to the assumption underlying our formulation of considering the energy quadratic in the polaron charge $q$ (cf.\ Sec.\ \ref{sec:semilocal}). Hence, the total energy of the polaron system can be expanded in a Taylor series as follows
\begin{align}
	E^0(q) & = E^0_\text{ref}(0) + q \left.\frac{dE^0(q)}{dq}\right|_0 + \frac{q^2}{2} \frac{d^2 E^0(q)}{dq^2} \nonumber \\
	& \quad + E^0(0) - E_\text{ref}^0(0)\label{eq:expansion_sio},
\end{align}
where $E^0_\text{ref}(0)$ is the PBE energy of the neutral undistorted system, $E^0(0)$ is the PBE energy of the neutral system with polaronic distortions, and the derivatives are taken at fixed polaron structure. Under the assumption that polaronic lattice distortions can be truncated at second order in the lattice displacements, the second line in Eq.\ \eqref{eq:expansion_sio} is written as
\begin{equation}
	E^0(0) - E_\text{ref}^0(0) = \frac 12 \sum_{\rho\rho'} C_{\rho\rho} \Delta \tau_{\rho} \Delta \tau_{\rho'}, \label{eq:cost_latt_sio}
\end{equation}
where $\rho=(i,\nu,p)$ denotes the Cartesian component $\nu$ of atom $i$ in the unit cell $p$, $C_{\rho \rho'}$ is the matrix of interatomic forces, and $\Delta \tau_{\rho}$ is the atomic displacement. In regard to the first and second order derivatives of the energy with respect to $q$ in Eq.\ \eqref{eq:expansion_sio}, electron screening effects accounting for the presence of the polaron charge are neglected \cite{sio2019PRB}. This implies that the valence wave functions are kept fixed to those obtained for the neutral state with polaronic lattice distortions. Thus, the first and second derivatives of the energy with respect to $q$ in Eq.\ \eqref{eq:expansion_sio} are replaced with their respective unscreened bare values, namely
\begin{align}
	\left.\frac{dE^0(q)}{dq}\right|_{0,\text{bare}} &= - \int\! d\mathbf r\, \psi_\text{p}^*(\mathbf r) \left.\frac{\delta E^0}{\delta n_{\sigma_\text{p}}(\mathbf r)}\right|_0  \psi_\text{p}(\mathbf r) , \label{eq:sio_1st_E} \\
	\left.\frac{d^2E^0(q)}{dq^2}\right|_\text{bare} &= \int\! d\mathbf r d\mathbf r' \frac{\delta^2 E^0}{\delta n_{\sigma_\text{p}}(\mathbf r) \delta n_{\sigma_\text{p}}(\mathbf r')} n_\text{p}(\mathbf r)n_\text{p}(\mathbf r'), \label{eq:sio_2nd_E}
\end{align}
where $\psi_\text{p}$ is the polaron wave function, $n_\text{p} = |\psi_\text{p}|^2$ the polaron density, where contributions due to the variations of the occupied valence wave functions are neglected, and $\sigma_\text{p}$  the spin channel hosting the polaron. Next, the functional derivative of the energy in Eq.\ \eqref{eq:sio_1st_E} is expanded to first order with respect to the lattice displacements, namely:
\begin{equation}
	\left.\frac{\delta E^0}{\delta n_{\sigma_\text{p}}}\right|_0 = \mathcal H^0_\text{ref}(0) + \sum_{\rho} \frac{dV^0_{\sigma_\text{p}}}{d\tau_{\rho}} \Delta \tau_{\rho}, \label{eq:ham0_exp}
\end{equation}
where $\mathcal H^0_\text{ref}(0)$ is the PBE Hamiltonian of the neutral undistorted reference system and $V_{\sigma_\text{p}}^0$ is the PBE Kohn-Sham potential for the spin channel $\sigma_\text{p}$. The expression in Eq.\ \eqref{eq:ham0_exp} describes  the Hamiltonian of the neutral charge state with polaronic lattice distortions.
The second order term in Eq.\ \eqref{eq:sio_2nd_E} is identified as a self-interaction term and the corresponding term  is removed from Eq.\ \eqref{eq:expansion_sio}. Hence, one obtains the following self-interaction corrected functional \cite{sio2019PRB}:
\begin{align}
	& E^\text{uc}(q)  = E^0_\text{ref}(0) + \frac 12 \sum_{\rho\rho'} C_{\rho \rho'} \Delta \tau_{\rho} \Delta \tau_{\rho'} \nonumber \\
	& \quad + q \int\! d\mathbf r\, \psi_\text{p}^*(\mathbf r) \left[ \mathcal H^0_\text{ref}(0) + \sum_{\rho} \frac{dV^0_{\sigma_\text{p}}}{d\tau_{\rho}} \Delta \tau_{\rho} \right]  \psi_\text{p}(\mathbf r). \label{eq:etot_sio_final}
\end{align}
This functional is minimized with respect to the polaron wave function and the polaronic lattice distortions. The resulting self-consistent equations are then reformulated by expanding the polaron wave function in the basis of Kohn-Sham states, and the polaronic distortions in the basis of phonon modes. Within this approach, the polaron formation energy is determined as \cite{sio2019PRB}:
\begin{equation}
	E_\text{f}^\text{uc} = q(\epsilon_\text{b}^0 - \epsilon_\text{p}^\text{uc}) + \frac 12 \sum_{\rho\rho'} C_{\rho\rho'} \Delta \tau_\rho \Delta \tau_{\rho'}, \label{eq:Ef_sio}
\end{equation}
where the polaron level $\epsilon_\text{p}^\text{uc}$ is obtained from the following Kohn-Sham equation:
\begin{equation}
	\Bigg( \mathcal H^0_\text{ref}(0) + \sum_{\rho} \frac{dV^0_{\sigma_\text{p}}}{d\tau_{\rho}} \Delta \tau_{\rho} \Bigg)  \psi_\text{p} = \epsilon_\text{p}^\text{uc} \psi_\text{p}. \label{eq:ks_sio}
\end{equation}
The  expression in Eq.\ \eqref{eq:Ef_sio} highlights the sum of two energy contributions due to charge localization and to lattice distortions, respectively. This carries the same physical interpretation as Eqs.\ \eqref{eq:Ef_2} and \eqref{eq:Ef_2_sl}, which are found when the piecewise-linearity condition is enforced. 

\section{Connection Between Charged and Neutral Density Functionals \label{sec:equivalence}}

\subsection{Connection between $\gamma$DFT, $\mu$DFT,  and pSIC \label{sec:connection_gmDFT_pSIC}}

% energy
We now compare $\gamma$DFT, $\mu$DFT, and pSIC functionals. As discussed in Sec.\ \ref{sec:semilocal}, within the assumptions outlined in Sec.\ \ref{sec:unified_hybrid}, $\gamma$DFT, $\mu$DFT, and pSIC lead to the same formal expression of the self-interaction-corrected energy. Indeed, through Janak's theorem \cite{janak1978PRB}, the self-interaction–corrected energy obtained with $\gamma$DFT or $\mu$DFT in Eq.\ \eqref{eq:E_PWL_semilocal} coincides with that obtained in Eq.\ \eqref{eq:esic_0_first}, which also coincides with that of pSIC. This shows that the polaron energetics corrected for the many-body self-interaction in $\gamma$DFT, $\mu$DFT, and pSIC can be evaluated using the energetics of the neutral defect with polaronic distortions. The slope of the total energy with respect to the polaron charge is given by the polaron level, which is independent of its occupation in the limit of infinite supercell. 

% forces
A main difference between $\gamma$DFT, $\mu$DFT, and pSIC lies in the way the atomic forces are determined. In pSIC, the forces required for achieving the polaronic distortions of the polaron state at $q=\pm1$ are determined through an expression evaluated at infinitesimal charge states around $q=0$ [cf.\ Eq.\ \eqref{eq:fSIC_ff}]. At variance, in $\gamma$DFT and $\mu$DFT, the forces are calculated through a  self-consistent  calculation at $q=\pm 1$. For an infinite supercell, these two ways of proceeding allow one to obtain the correct atomic forces in view of the linear dependence of the considered functional on $q$  [cf.\ Eq.\ \eqref{eq:forces_sic}].  The validity of this equivalence holds provided the effect of higher-order dependencies of the total energy on the charge $q$ associated with the different functional forms of $\gamma$DFT, $\mu$DFT, and pSIC can be neglected.

% orbital dependence
Another difference between $\gamma$DFT, $\mu$DFT, and pSIC lies in the way the self-interaction potential correction is constructed. In particular, $\gamma$DFT and $\mu$DFT require the identification of a specific state for which the self-interaction needs to be corrected. In a convention where electrons are added to the spin up channel and removed from the spin down channel, the electron polaron state corresponds to the highest occupied state in the spin up channel, while the hole polaron state corresponds to the lowest unoccupied state in the spin down channel. The densities of these orbitals are then used to construct the localizing potentials in Eqs.\ \eqref{eq:vgamma} and \eqref{eq:vmu}. In contrast, in pSIC, the potential correction in Eq.\ \eqref{eq:mbSIC_pot_sl} does not require determining the density of a single individual orbital. This allows the self-interaction correction to be performed on a given charge state without the need to relate it to a specific orbital.

\subsection{Connection between the unit-cell method and pSIC \label{sec:connection_pSIC_unitcell}}

% alternative way of writing the Giustino's functional
In this section, we focus on the comparison between the unit-cell method for polarons introduced by Sio \textit{et al.}\ \cite{sio2019PRB,sio2019PRL} and pSIC. We first compare the treatment of the polaronic lattice distortions. In the unit-cell method, the energy cost due to lattice distortions is determined by assuming a quadratic expansion of the energy with respect to atomic displacements [cf.\ Eq.\ \eqref{eq:cost_latt_sio}]. Thus, higher-order contributions in the lattice displacements and the corresponding  contributions to the cost of lattice distortions in the Kohn-Sham equations \eqref{eq:ks_sio}, 
\begin{equation}
	\left( \sum_{\rho \rho'} \frac{d^2 V^0_{\sigma_\text{p}}}{d\tau_\rho d\tau_{\rho'}} \Delta \tau_\rho \Delta \tau_{\rho'} + \dots \right) \psi_\text{p},
\end{equation}
are neglected in the unit-cell method. Anharmonic effects in phonon scattering phenomena are crucial when modeling, for instance, transport properties in disordered or anharmonic systems \cite{simoncelli2023npj}. Such effects might also be critical for polarons in liquids, such as the hydrated electron in water \cite{lan2021simulating,lan2022angewandte,lan2024dynamics}. Hence, the applicability of the unit-cell method to these cases remains to be ascertained. In contrast, pSIC does not require any assumptions about polaronic distortions, and the cost due to lattice distortions is calculated by taking the difference between the total energies of the neutral distorted and neutral undistorted systems, without resorting to any approximation. This approach allows pSIC to account for anharmonic contributions in the cost of lattice distortions, and is generally applicable to anharmonic systems, disordered systems, or liquids.

% Comparison with pSIC: addition of the charge
Next, we compare the unit-cell method and pSIC regarding the contributions to the energy of the neutral system arising from the addition of the polaron charge. In the unit-cell method, these contributions are the first-order and second-order derivatives of the total energy with respect to the polaron charge $q$. Then, screening effects are neglected, resulting in the expressions in Eqs.\ \eqref{eq:sio_1st_E} and \eqref{eq:sio_2nd_E}. The first-order derivative in Eq.\ \eqref{eq:sio_1st_E} can be rewritten as
\begin{equation}
	\left.\frac{dE^0(q)}{dq}\right|_{0,\text{bare}} = -\braket{\psi_\text{p} |\mathcal H^0_{\sigma_\text{p}}(0) | \psi_\text{p}} = -\epsilon_\text{p}^0(0),
\end{equation}
which coincides with the opposite of the energy level of the neutral polaron state $\epsilon_\text{p}^0(0)$. The second-order derivative in Eq.\ \eqref{eq:sio_2nd_E} is identified as the self-interaction term and is thus removed. This corresponds to defining the following self-interaction energy correction
\begin{widetext}
	\begin{equation}
		\left.\Delta E^{0}(q)\right|_\text{bare} = - \frac{q^2}{2}\left. \frac{d^2 E^0(q)}{dq^2}\right|_\text{bare}  = - q^2   \Bigg\lbrace E_\text{H}[n_\text{p}]  + \frac{1}{2} \!\int\!\! d\mathbf r d\mathbf r'\, \frac{\delta^2 E_\text{xc}[n_\uparrow,n_\downarrow]}{\delta n_{\sigma_\text{p}}(\mathbf r) \delta n_{\sigma_\text{p}}(\mathbf r')} n_\text{p}(\mathbf r) n_\text{p}(\mathbf r')   \Bigg\rbrace, \label{eq:Esic_bare}
	\end{equation}
\end{widetext}
which coincides with the expression given in Ref.~\cite{sio2019PRB}, once the neutralizing background charge in the Hartree term is taken into account. We note that, compared to Eq.\ \eqref{eq:mbSIE_sl}, the expression in Eq.\ \eqref{eq:Esic_bare} replaces the derivatives $dn_\sigma/dq$ of the total density with respect to the polaron charge by the polaron density $n_\text{p}$, thereby neglecting the first-order variations of the valence electron density upon polaron addition. For this reason, we regard it as a bare self-interaction energy. At variance, in pSIC, screening effects  are fully accounted for in the definition of the self-interaction energy correction and are not neglected. The contributions arising from the addition of the charge are written as first-order and second-order derivatives of the total energy with respect to $q$, similarly to the unit-cell method. Through Janak's theorem \cite{janak1978PRB}, the first-order derivative of the total energy with respect to $q$  is expressed as the energy level of the neutral state with polaronic distortions, namely
\begin{equation}
	\left.\frac{dE^0(q)}{dq}\right|_0 = - \epsilon_\text{p}^0(0). \label{eq:janak_neutraleps}
\end{equation}
The second-order derivative of the total energy with respect to $q$ is regarded as a many-body self-interaction [cf.\ Eq.\ \eqref{eq:mbSIE_sl}] and is removed from the total energy. 

% direct comparison
In comparing the unit-cell method and pSIC, we remark that the right hand sides of Eqs.\ \eqref{eq:sio_1st_E} and \eqref{eq:janak_neutraleps} coincide, despite the extra assumption of neglecting screening effects when deriving Eq.\ \eqref{eq:sio_1st_E} in the unit-cell method. 
This is due to Janak's theorem \cite{janak1978PRB} and does not require neglecting screening effects. This is also related to the fact that changes in the valence wave functions do not contribute to the total energy in first-order perturbation. In regard to the self-interaction terms in Eqs.\ \eqref{eq:mbSIE_sl}  and \eqref{eq:Esic_bare}, we note that the many-body self-interaction energy reduces to the bare self-interaction energy when screening effects are suppressed, namely
\begin{align}
	\Big.\left.\Delta E^{0}(q)\right|_\text{mb}^\text{sl}\Big|_\text{no screening} = \left.\Delta E^{0}(q)\right|_\text{bare}. 
\end{align}
Hence, suppressing screening effects and removing the bare self-interaction, as done by Sio \textit{et al.}, leads to the same functional form of the self-interaction corrected energy as obtained by accounting for screening effects and removing the many-body self-interaction, as done in pSIC. As a result, the energy gain due to charge localization obtained with the unit-cell method and  pSIC coincide for an harmonic system. For an anharmonic system, more pronounced differences could arise between the two methods. In this case, the polaron level calculated with the unit-cell method lacks higher-order contributions to the Hamiltonian in Eq.\ \eqref{eq:ks_sio}, which are instead accounted for in pSIC. 

Finally, we compare the unit-cell method and pSIC in relation to the way atomic forces are determined, which affects the resulting polaronic distortions. In pSIC, forces are determined through an expression evaluated at infinitesimal charge states around $q=0$ [cf.\ Eq.\ \eqref{eq:fSIC_ff}], which explicitly includes electron screening effects due to the addition of the extra charge. At variance, the wave functions obtained with the unit-cell method when minimizing Eq.\ \eqref{eq:ks_sio} do not account for electron screening effects, as the Hamiltonian in  Eq.\ \eqref{eq:ks_sio} corresponds to that of the neutral state with polaronic distortions. As a result, the polaronic lattice distortions obtained with the unit-cell method lack contributions from electron screening effects, which are instead accounted for in pSIC. It is expected that this difference leads to atomic forces of different size and, in principle, to different equilibrium geometries for polaronic states.

% DFT sic functional by sio et al
In addition to introducing the unit-cell method for polarons, Ref.\ \cite{sio2019PRB} proposes a charged supercell approach, whose formulation is inspired by that of the unit-cell method. In particular, the charged supercell method in Ref.\ \cite{sio2019PRB} is obtained by adding the self-interaction energy correction in Eq.\ \eqref{eq:Esic_bare} to the total energy functional, together with the contribution from the neutralizing background charge in the Hartree term. The corresponding self-interaction–corrected Kohn–Sham equations are then derived by variational differentiation, yielding a Kohn–Sham Hamiltonian that acts differently on the polaron state and on the occupied valence states. Specifically, the Hamiltonian acting on the valence electrons includes corrections only to the exchange–correlation potential, whereas that acting on the polaron state includes corrections to both the Hartree and the exchange–correlation potentials. In this self-consistent procedure, both the valence and polaron wave functions are variationally optimized in the presence of the polaron charge. As discussed in Sec.\ \ref{sec:connection_pSIC_unitcell}, the energy correction in Eq.\ \eqref{eq:Esic_bare} does not account for variations in the valence electron density with respect to the polaron charge, unlike the energy correction in Eq.\ \eqref{eq:mbSIE_sl}. Consequently, although the valence wave functions are self-consistently optimized in the presence of the polaron charge in the supercell method in Ref.\ \cite{sio2019PRB}, the resulting energy corrections were found to give overestimated formation energies \cite{falletta2022PRB,falletta2023thesis}. The overestimated polaron formation energies reported in Refs.~\cite{falletta2022PRB,falletta2023thesis} were obtained by evaluating the self-interaction-corrected energetics proposed by Sio \textit{et al.}\ using localized polaron wave functions calculated with hybrid functionals. This overestimation arises because the absence of electron screening effects in the definition of the self-interaction energy correction in Eq.\ \eqref{eq:Esic_bare} leads to an overstabilization of the polaron state when solving the Kohn–Sham equations. This overestimation is especially pronounced for small polarons, where electron screening effects are sizeable \cite{falletta2022PRL,falletta2022PRB}, and is expected to be less significant for large polarons.  To summarize, the unit-cell method in Ref.\ \cite{sio2019PRB} retrieves the many-body self-interaction–corrected energetics similarly to the pSIC method, whereas the charged supercell method in Ref.\ \cite{sio2019PRB} tends to overestimate polaron formation energies due to the lack of electron screening effects in the self-interaction energy correction.

\section{Finite-size corrections for charged and neutral polaronic states \label{sec:finite-size}}

\begin{figure*}[t!]
	\includegraphics[width=\linewidth]{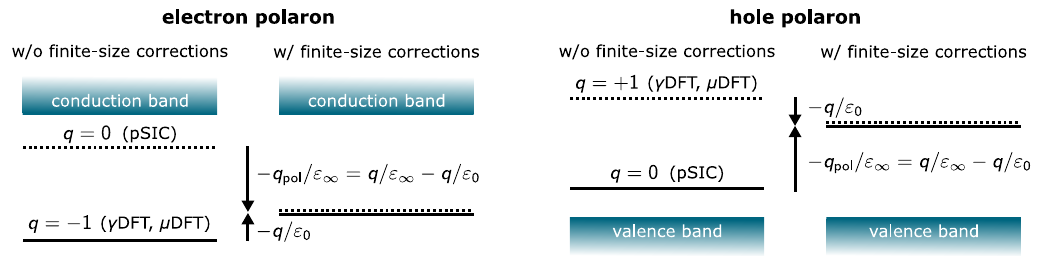}
	\caption{Different treatment of finite-size effects on polaron energy levels when using $\gamma$DFT at $\gamma_\text{k}$, $\mu$DFT at $\mu_\text{k}$, and pSIC for electron and hole polarons in a finite-size supercell. The finite-size corrections on the polaron energy levels are given in Eqs.\ \eqref{eq:eps_fs_integerq} and \eqref{eq:eps_fs_zeroq}. Occupied levels are given by solid lines, unoccupied levels by dashed lines. Without applying finite-size corrections, the energy gain due to charge localization is underestimated in pSIC and overestimated in $\gamma$DFT and $\mu$DFT. This may lead to underdistorted polaronic structures in pSIC, and overdistorted polaronic structures in $\gamma$DFT and $\mu$DFT. These differences are reduced when increasing the size of the supercell.}
	\label{fig:finite-size}
\end{figure*}

% recap on finite-size corrections
When modeling polarons in supercells, one needs to account for finite-size effects arising from the Coulomb interactions of the polaronic defect with its periodic images and with the neutral background charge \cite{freysoldt2009PRL,komsa2012PRB,kumagai2014PRB,falletta2020PRB,walsh2021npj}. These interactions affect both the charged and the neutral polaronic states. Finite-size effects can occur also in the neutral state  because of the presence of ionic polarization charges associated with the polaronic distortions \cite{falletta2020PRB}. The finite-size correction for the total energy of a calculation with charge $q'$ localized at a structure $\mathbf R_q$, which has been obtained through relaxation in the presence of a charge $q$, is given by \cite{falletta2020PRB}
\begin{align}
	E_\text{cor}(q',\mathbf R_{q}) & = E_\text{m}(q,\varepsilon_0) - E_\text{m}(q + q_\text{pol},\varepsilon_\infty) \nonumber \\
	& \quad + E_\text{m}(q'+q_\text{pol},\varepsilon_\infty), \label{eq:Etot_fs}
\end{align}
where $E_\text{m}(q,\varepsilon)$ is the correction to the total energy of a system of charge $q$ and with screening described by a dielectric constant $\varepsilon$,  as calculated with the model of Freysoldt, Neugebauer, and Van de Walle \cite{freysoldt2009PRL}, $\varepsilon_\infty$ is the high-frequency dielectric constant, $\varepsilon_0$ is the static dielectric constant, and
\begin{equation}
	q_\text{pol}=-q\Big(1-\frac{\varepsilon_\infty}{\varepsilon_0} \Big) \label{eq:qpol}
\end{equation}
is the ionic polarization charge associated with the polaronic lattice distortions \cite{falletta2020PRB}. The finite-size corrections $E_\text{m}(q,\varepsilon)$ are quadratic in the charge $q$ and inversely proportional to the dielectric constant $\varepsilon$. 
For vertical transition energies, it can be shown \cite{falletta2020PRB} that Eq.\ \eqref{eq:Etot_fs} yields equivalent corrections to those given in Ref.\ \cite{gake2020PRB}. 

Applying Janak's theorem \cite{janak1978PRB} to Eq.\ \eqref{eq:Etot_fs}, one derives the corresponding correction for the polaron level \cite{falletta2020PRB}, namely
\begin{equation}
	\epsilon_\text{cor}(q',\mathbf R_{q}) = - 2 \frac{E_\text{m}(q'+q_\text{pol},\varepsilon_\infty)}{q'+q_\text{pol}}. \label{eq:eps_fs}
\end{equation}
The code for calculating these corrections is freely available  \cite{gitfolder}. The addition of such finite-size corrections to the energy and polarons levels is implicitly assumed in the notation of all formulas in this work. 

% specialization in the case of polaron states
We now specialize Eqs.\ \eqref{eq:Etot_fs} and \eqref{eq:eps_fs} to the relevant polaronic states needed to enforce the piecewise-linearity condition in Eq.\ \eqref{eq:PWL}. In the case of a polaron state at integer charge, $q = q'$,  with  $q=-1$ for electron polarons and $q=+1$ for hole polarons. Hence, Eqs.\ \eqref{eq:Etot_fs} and \eqref{eq:eps_fs} reduce to the expressions in Refs.\ \cite{freysoldt2009PRL} and \cite{chen2013PRB}, respectively, namely
\begin{align}
	E_\text{cor}(q,\mathbf R_{q}) & = E_\text{m}(q,\varepsilon_0), \label{eq:Etot_fs_integerq} \\
	\epsilon_\text{cor}(q,\mathbf R_{q}) & = - 2 \frac{E_\text{m} (q,\varepsilon_0)}{q}, \label{eq:eps_fs_integerq}
\end{align}
where the minus sign in Eq.\ \eqref{eq:eps_fs_integerq} derives from the use of Janak's theorem \cite{janak1978PRB}, i.e., $\epsilon_\text{cor}(q,\mathbf R_q) = -d E_\text{cor}(q,\mathbf R_q) / d q$. However, different expressions need to be used for a polaron state in the neutral charge state $q' = 0$, when the underlying structure corresponds to the relaxed structure of an electron ($q= -1$) or a hole ($q=  +1$)  polaron. In this case, using the fact that $E_\text{m}(q,\varepsilon)$ is quadratically dependent on $q$ and inversely proportional to $\varepsilon$, it is straightforward to see that Eqs.\ \eqref{eq:Etot_fs} and \eqref{eq:eps_fs} reduce to:
\begin{align}
	E_\text{cor}(0,\mathbf R_{q}) & = E_\text{m}(q,\kappa), \label{eq:Etot_fs_zeroq} \\
	\epsilon_\text{cor}(0,\mathbf R_{q}) & = 2 \frac{E_\text{m} (q,\kappa)}{q}, \label{eq:eps_fs_zeroq}
\end{align}
where $1/\kappa = 1/\varepsilon_\infty - 1/\varepsilon_0$. 

We note that the finite-size correction to the charged polaron level is proportional to $-q/\varepsilon_0$, which corresponds to the polaron charge $q$, electronically and ionically screened through the static dielectric constant $\varepsilon_0$. At variance, the finite-size correction to the neutral polaron level is proportional to $q/\kappa$, which can be rewritten as
\begin{equation}
	\frac{q}{\kappa} = q \Big( \frac{1}{\varepsilon_\infty} - \frac{1}{\varepsilon_0} \Big) = -\frac{q_\text{pol}}{\varepsilon_\infty}, \label{eq:qpol_kappa}
\end{equation}
where we use the expression of $q_\text{pol}$ given in Eq.\ \eqref{eq:qpol}.  Equation \eqref{eq:qpol_kappa} highlights that the finite-size correction on the neutral polaron level is proportional to the ionic polarization charge due to lattice distortions, electronically screened through the high-frequency dielectric constant $\varepsilon_\infty$. 

We remark that the corrections for the charged and neutral polaron levels are opposite in sign, due to the opposite sign of $q$ and $q_\text{pol}$ [cf.\ Eq.\ \eqref{eq:qpol}]. In particular, for electron polarons,  $\epsilon_\text{cor}(-1,\mathbf R_{-1})$ is positive and  $\epsilon_\text{cor}(0,\mathbf R_{-1})$ is negative. At variance, for hole polarons,  $\epsilon_\text{cor}(+1,\mathbf R_{+1})$ is negative and $\epsilon_\text{cor}(0,\mathbf R_{+1})$ is positive. This indicates that the use of finite supercells overstabilizes the charged polaron state and destabilizes the polaron state at zero charge. The destabilization of the neutral charge state is more pronounced than the stabilization of the charged state, since $\kappa < \varepsilon_0$.

%Finite-size effects
In the case of supercell methods, finite-size effects are generally treated \emph{a posteriori}, leading to significant differences between $\gamma$DFT, $\mu$DFT, and pSIC. Since such finite-size corrections are not self-consistently applied to the Kohn-Sham equations, 
the forces resulting from polaron localization are not corrected for finite-size effects. The finite-size effects of the neutral defect state ($q=0$) with polaronic distortions are significant, as they are proportional to  $q/\kappa = q/\varepsilon_\infty - q/\varepsilon_0 = -q_\text{pol}/\varepsilon_\infty$ [cf.\ Eq.\ \eqref{eq:qpol_kappa}], where $q_\text{pol}$ is the ionic polarization charge due to the presence of polaronic distortions, as defined in Eq.\ \eqref{eq:qpol}. In addition to being sizeable, these effects on the neutral polaron state favor the delocalization of the polaron wave function. Indeed, they cause the neutral defect level to be energetically closer to the delocalized band-edge states, thereby reducing the energy gain due to charge localization. \emph{A posteriori} finite-size correction schemes are then used to correct the energy to achieve the limit of the infinite supercell. However, disregarding finite-size corrections on the forces results in less distorted polaron structures due to the underestimation of the energy gain associated with charge localization. This can lead to polaron delocalization in small supercells during structural relaxations, especially when $q/\kappa = -q_\text{pol}/\varepsilon_\infty$ is large. As pointed out by Sio \textit{et al.}\ \cite{sio2019PRB}, this phenomenon is analogous to the metal-insulator Mott transition, where spurious overlap of the polaron wave functions with its periodic images forms an extended wave function preventing polaron localization. In the case of $\gamma$DFT and $\mu$DFT, finite-size effects on the forces need to be addressed too. However, compared to the finite-size corrections on the neutral state with polaronic distortions, the finite-size effects on the charged polaron state are less significant and carry the opposite sign, as they are proportional to $-q/\varepsilon_0$ instead of $q/\kappa = -q_\text{pol}/\varepsilon_\infty$. For these reasons, $\gamma$DFT and $\mu$DFT tend to overlocalize polarons in finite-size supercells and are more likely to localize polarons in very small supercells compared to the neutral pSIC scheme. These spurious finite-size effects on forces are expected to be marginal for systems with large dielectric constants. A schematic outline of finite-size effects in $\gamma$DFT, $\mu$DFT, and pSIC is given in Fig.\ \ref{fig:finite-size}.

In the case of the unit-cell method, finite-size effects are generally adressed through the use of increasingly dense  $\mathbf k$-point samplings \cite{sio2019PRB,sio2019PRL}. The final results are then obtained through extrapolation to the dilute limit.

\section{Results \label{sec:results}}

\begin{figure*}[t!]
	\includegraphics[width=\linewidth]{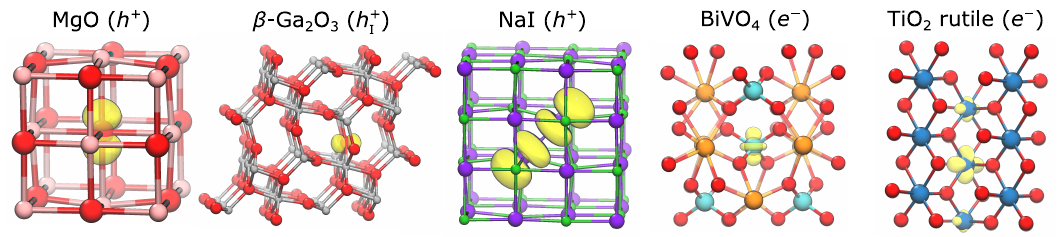}
	\caption{Polaron isodensity surface at 5\% of its maximum calculated at $q=0$ for the hole polaron in MgO (216-atom supercell), the hole polaron in $\beta$-Ga$_2$O$_3$ (120-atom supercell), the $V_\text{k}$ center in NaI (216-atom supercell), the electron polaron in BiVO$_4$ (96-atom supercell), and the electron polaron in TiO$_2$ (216-atom supercell). Mg in pink, O in red, Ga in grey, Na in green, I in violet, Bi in orange, V in cyan, and Ti in blue. The vertical axis is the $z$ axis.}
	\label{fig:polarons}
\end{figure*}

\subsection{Computational details \label{sec:comput_details}}

% computational details
The DFT calculations are performed using the \textsc{quantum espresso} suite \cite{giannozzi2009JPCM,giannozzi2017JPCM}. Version 7.2 includes the implementations of the $\gamma$DFT functional and of the self-consistent scissor operator. We will make the implementations of the $\mu$DFT and pSIC functionals available for incorporation in an upcoming official release of \textsc{quantum espresso}. The core-valence interactions are represented using norm-conserving pseudopotentials \cite{vansetten2018CPC}. We study the hole polaron in MgO \cite{varley2012PRB,kokott2018NJP,wing2020PRM,falletta2022PRL,falletta2022PRB}, the hole polaron in  $\beta$-Ga$_2$O$_3$  \cite{varley2012PRB,deak2017PRB,ho2018PRB,gake2019PRM,frodason2020JAP,bouzid2019PSS,falletta2023PRB}, the $V_\text{k}$ center in NaI \cite{sadigh2015PRB,miceli2018PRB}, the electron polaron in monoclinic BiVO$_4$  \cite{wiktorACS2018,pham2020JCTC,falletta2022PRL,falletta2022PRB},  and the electron polaron in rutile TiO$_2$ \cite{deskins2007PRB,kokott2018NJP,pham2020JCTC,reticcioli2022JACM}. In  $\beta$-Ga$_2$O$_3$, there are three differently coordinated O atoms, and correspondingly three different hole polarons. We here consider the hole polaron localized on the O$_\text{I}$ site, which is shared by two GaO$_6$ octahedra and one GaO$_4$ tetrahedron. We illustrate the isosurfaces of the considered polarons in Fig.\ \ref{fig:polarons}. The lattice parameters are calculated at the PBE level of theory for the pristine systems. The Brillouin zone sampling is conducted at the $\Gamma$ point. An energy cutoff of 100 Ry is used for all calculations. MgO is modeled with a 216-atom 3$\times$3$\times$3 cubic supercell ($a = 4.22$ \AA), $\beta$-Ga$_2$O$_3$ with a 120-atom 1$\times$3$\times$2 monoclinic supercell ($a=12.38$ \AA{}, $b=3.09$ \AA{}, $c=5.88$ \AA{}, $\beta=103.82^\circ$), NaI with a 216-atom 3$\times$3$\times$3 cubic supercell ($a = 6.52$ \AA), BiVO$_4$ with a 96-atom 2$\times$2$\times$2 orthorhombic supercell ($a = 5.17$ \AA, $b=5.17$ \AA, $c = 5.90$ \AA), and rutile TiO$_2$ with a 216-atom 3$\times$3$\times$4 orthorhombic supercell ($a = 4.64$ \AA, $c = 2.98$ \AA). These values are in agreement with the respective experimental lattice parameters of MgO ($a^\text{expt}=4.207$ \AA{}  \cite{smith1968JAC}), $\beta$-Ga$_2$O$_3$ ($a^\text{expt}=12.214$ \AA{}, $b^\text{expt}=3.037$ \AA{}, $c^\text{expt}=5.798$ \AA{}, $\beta=103.83^\circ$ \cite{aahman1996ACSC}), NaI ($a^\text{expt}=6.47$ \AA{} \cite{posnjak1922JWAS}), BiVO$_4$ ($a^\text{expt} = 5.194$ \AA{}, $b^\text{expt} = 5.090$ \AA{}, $c^\text{expt}=5.849$ \AA{} \cite{sleight1979MRB}),  and  TiO$_2$ ($a^\text{expt}=4.593$ \AA{}, $c^\text{expt}=2.958$ \AA{} \cite{burdett1987JACS}). By applying finite electric fields \cite{umari2002PRL} at the PBE level of theory, we determine the high-frequency and static dielectric constants. The calculated average dielectric constants $\varepsilon = (\varepsilon_{xx} + \varepsilon_{yy} + \varepsilon_{zz})/3$  are given in Table \ref{tab:dielectric_constants}. For the pSIC calculations, we use $|\delta q| = 0.01$ a.u., which is found to be large enough to avoid numerical instabilities, and small enough to converge the finite difference expressions of  the forces. The calculations with the unit-cell method are carried out with the code EPW \cite{noffsinger2010CPC,lee2023npj}. For MgO and NaI, we use homogeneous $12\times12\times12$ meshes for $\mathbf k$-point and $\mathbf q$-point samplings for calculating the electronic bands and the electron-phonon couplings, respectively. For $\beta$-Ga$_2$O$_3$, we use a $10\times10\times12$ mesh for $\mathbf k$-point sampling and a $5\times5\times6$ for $\mathbf q$-point sampling. For BiVO$_4$,  we use a $8\times8\times8$ mesh for $\mathbf k$-point sampling and a $4\times4\times4$ for $\mathbf q$-point sampling. For TiO$_2$, we use a $12\times12\times16$ mesh for $\mathbf k$-point sampling and a $6\times6\times8$ for $\mathbf q$-point sampling. An energy cutoff of 100 Ry is used in all cases. For TiO$_2$, we use the PBEsol functional \cite{perdew2008PRL}. The $\mathbf k$ path along high symmetry points was generated with SeeK-path \cite{hinuma2017CMS}. For the band structure calculation, we interpolate  between each two consecutive high-symmetry $\mathbf k$-points using 100 points. The Wannierization is carried out on the O $2p$ orbitals for MgO and $\beta$-Ga$_2$O$_3$, on the I $2p$ orbitals for NaI, on the V $3d$ orbitals for BiVO$_4$, and on the Ti $3d$ orbitals for TiO$_2$.

\begin{table}[h]
        \caption{High-frequency dielectric constant $\varepsilon_\infty$, static dielectric constant $\varepsilon_0$, strength $\gamma_\text{k}$ of the $\gamma$DFT functional, and strength $\mu_\text{k}$ of the $\mu$DFT functional. The parameters $\gamma_\text{k}$ and $\mu_\text{k}$ enforce the piecewise linearity of the total energy [cf.\ Eq.\ \eqref{eq:PWL}] after including finite-size corrections (cf.\ Sec.\ \ref{sec:finite-size}).}
	\label{tab:dielectric_constants}
	\begin{ruledtabular}
		\begin{tabular}{l *{4}{c} }	
			System    &  $\varepsilon_\infty$ & $\varepsilon_0$ & $\gamma_\text{k}$ & $\mu_\text{k}$  \\  \hline \rule{-4pt}{3ex}
                MgO ($h^+$)                    &  2.95  &  10.70  & 1.87  & 10.39 \\  \rule{-4pt}{3ex}
                $\beta$-Ga$_2$O$_3$ ($h_\text{I}^+$)   &  3.75  &  11.98  & 1.38  & 3.45   \\  \rule{-4pt}{3ex} 
                NaI ($h^+$)                    &  3.12  &  8.22   & 3.67  & 48.06 \\  \rule{-4pt}{3ex}
			BiVO$_4$ ($e^-$)               &  5.83  &  64.95  & 1.80  & 2.85  \\  \rule{-4pt}{3ex} 
                TiO$_2$ ($e^-$)                &  6.36  &  111.88  & 1.57  & 2.94   \rule{-4pt}{3ex} 
		\end{tabular}
	\end{ruledtabular}
\end{table}	

\subsection{Comparison between $\gamma$DFT, $\mu$DFT, and pSIC}

\begin{figure}[t!]
    \includegraphics[width=1.0\linewidth]{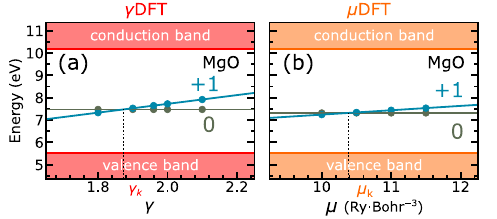}
    \caption{Charged and neutral polaron energy levels as a function of (a) $\gamma$ in $\gamma$DFT calculations and (b) $\mu$ in $\mu$DFT calculations for the hole polaron in MgO (216-atom supercell). The polaron levels are identified by their respective polaron charge states, namely +1 for the charged and 0 for the neutral state. The geometry is kept fixed at that obtained for $\gamma = \gamma_\text{k}$ and $\mu =  \mu_\text{k}$, respectively.}
    \label{fig:xi_k}
\end{figure}

% optimal parameter
We compare $\gamma$DFT, $\mu$DFT, and pSIC for the set of polarons considered in this work. First, we initialize the structure by distorting the crystal symmetry to induce the polaronic structure. In the case of polarons localized on a single site (MgO, $\beta$-Ga$_2$O$_3$, BiVO$_4$, TiO$_2$), this can be achieved by elongating the lattice bonds around the site where the polaron localizes (O for the hole polaron in MgO, O for the hole polaron in $\beta$-Ga$_2$O$_3$, V for the electron polaron in BiVO$_4$, Ti for the electron polaron in TiO$_2$). In the case of polarons localized on two sites (NaI), the atoms where the polaron localizes are put closer to each other (neighboring I atoms for the $V_\text{k}$ center in NaI). Then, we perform electronic and structural relaxations to find the optimized polaron structure. In pSIC, functional tuning is not required. In $\gamma$DFT and $\mu$DFT, one needs to find the optimal values of $\gamma_\text{k}$ and $\mu_\text{k}$ such that the piecewise-linearity condition in Eq.\ \eqref{eq:PWL} is ensured. As in Sec.\ \ref{sec:nonempirical_tuning}, we use  $\xi$ to commonly denote $\gamma$ and $\mu$. The parameter $\xi_\text{k}$ is determined by first performing structural relaxations for various values of $\xi$, then plotting the energy levels of both charged and neutral polarons in the presence of lattice distortions for each $\xi$, and finally identifying the value of $\xi$ where the lines interpolating the charged and neutral polaron levels intersect. This procedure is described in Sec.\ \ref{sec:nonempirical_tuning} and illustrated in Fig.\ \ref{fig:xi_k} in the case of the hole polaron in MgO. In Table \ref{tab:dielectric_constants}, we give the values of $\gamma_\text{k}$ and $\mu_\text{k}$ for the various polarons considered in this work.

\begin{table}[t]
    \caption{Lengths (in \AA) of polaronic lattice bonds for the hole polaron in MgO (O--Mg), the hole polaron in  $\beta$-Ga$_2$O$_3$ (O--Ga), the $V_\text{k}$ center in NaI (I--I), the electron polaron in BiVO$_4$ (V--O), and the electron polaron in rutile TiO$_2$ (Ti--O), as obtained with $\gamma$DFT, $\mu$DFT, and pSIC. For a given polaron, bond lengths are given in increasing order.} 
	\label{tab:structures}
	\begin{ruledtabular}
		\begin{tabular}{l *{3}{c} }	
		    System                                 & $\gamma$DFT         & $\mu$DFT       & pSIC  \\  \hline \rule{-4pt}{3ex} \rule{-4pt}{3ex}
MgO ($h^+$)                            & 2.21/2.32           & 2.21/2.32      & 2.18/2.25               \\ \rule{-4pt}{3ex} 
$\beta$-Ga$_2$O$_3$ ($h_\text{I}^+$)   & 2.02/2.16/2.16      & 2.00/2.16/2.16 & 1.95/2.13/2.13          \\ \rule{-4pt}{3ex} 
NaI ($h^+$)                            & 2.88                & 3.17           & 3.37                    \\ \rule{-4pt}{3ex} 
BiVO$_4$ ($e^-$)                       & 1.82                & 1.82           & 1.79                    \\ \rule{-4pt}{3ex} 
TiO$_2$ ($e^-$)                        & 2.00/2.06/2.07      & 2.01/2.06/2.08 & 2.00/2.02/2.03          \rule{-4pt}{3ex} 
		\end{tabular}
	\end{ruledtabular}
\end{table}	

\begin{figure*}[t!]
    \includegraphics[width=1.0\linewidth]{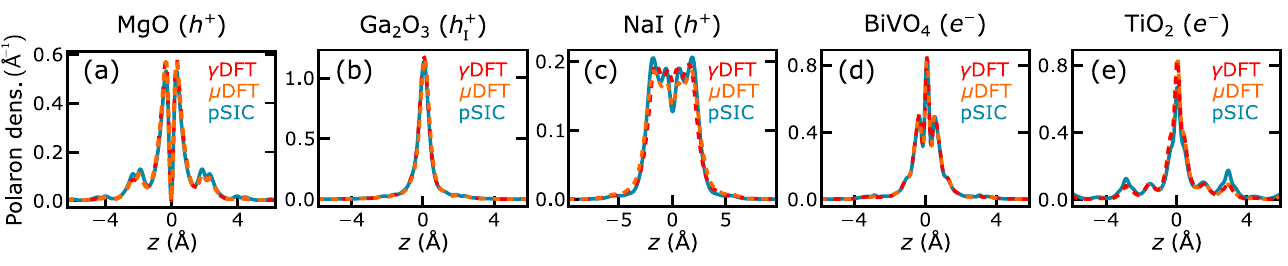}
    \caption{Polaron densities integrated over $xy$ planes obtained with $\gamma$DFT, $\mu$DFT, and pSIC for the various polarons considered in this work. The polaron densities are calculated for the neutral charge state $(q=0)$ in the presence of polaronic distortions, and are averaged over planes according to Eq.\ \eqref{eq:average_density}.}
    \label{fig:densities}
\end{figure*}

% comparison structures
We now compare the polaron structures obtained with piecewise-linear $\gamma$DFT, $\mu$DFT, and pSIC functionals. In MgO, we find four elongated O--Mg bonds lying in a plane and two longer O--Mg bonds out of the plane. In $\beta$-Ga$_2$O$_3$, the O--Ga bonds are elongated. In NaI, we find the $V_\text{k}$ center. In BiVO$_4$, the V--O bonds are all equally elongated. In TiO$_2$, the Ti--O are elongated. The lattice bonds pertaining to all polarons studied in this work are given in Table \ref{tab:structures}. Overall, we find excellent agreement for the polaronic distortions obtained with  $\gamma$DFT, $\mu$DFT, and pSIC. Specifically, discrepancies only amount to at most 0.07 \AA{}, excluding the case of the $V_\text{k}$ center in NaI. This demonstrates that piecewise-linear functionals are expected to yield polaronic distortions in close agreement to each other. For the I--I bond in NaI, we find a difference of 0.20 \AA{} between the $\mu$DFT  and pSIC results, and a larger difference of 0.29 \AA{} between the $\mu$DFT and $\gamma$DFT results. This is due to the weak nature of the I--I bond in NaI \cite{sadigh2015PRB}. With the pSIC scheme implemented in Ref.\ \cite{sadigh2015PRB}, the I--I bond was found to be equal to 3.38  \AA{}, in excellent agreement with our present pSIC result, with the minor difference of 0.01 \AA{} due to the use of a different code. 

% finite size effects
We note that pSIC leads to a less distorted structure compared to $\gamma$DFT or $\mu$DFT. As discussed in detail in Sec.\ \ref{sec:connection_gmDFT_pSIC} and illustrated in Fig.\ \ref{fig:finite-size}, this is due to the fact that, for a finite supercell,  pSIC is expected to underestimate the polaronic distortions while $\gamma$DFT and $\mu$DFT are expected to overestimate the polaronic distortions.  Indeed,  the energy level of the neutral polaron state obtained with pSIC before applying the finite-size corrections underestimates the energy gain due to charge localization. At variance, the charged polaron level obtained with $\gamma$DFT and $\mu$DFT before applying the finite-size corrections  overestimates the energy gain due to charge localization. Atomic forces obtained with  pSIC, $\gamma$DFT, and $\mu$DFT do not include any treatment of finite-size effects, as these are applied fully \emph{a posteriori}. As an  example of underestimated polaronic distortions in pSIC, we indicate the case of the hole polaron in a 64-atom supercell of MgO, which results in a  delocalized state in  pSIC, while $\gamma$DFT and $\mu$DFT lead to a localized state  \cite{falletta2022PRL,falletta2022PRB,falletta2024JAP}. 

% discrepancies for an infinite system
As discussed in Sec.\ \ref{sec:connection_pSIC_unitcell}, additional discrepancies between $\gamma$DFT, $\mu$DFT, and pSIC can stem from their distinct determination of atomic forces. In $\gamma$DFT and $\mu$DFT, forces are evaluated using the Hellmann–Feynman theorem after introducing the Hamiltonian corrections in Eqs.\ \eqref{eq:vgamma} and \eqref{eq:vmu}, respectively, whereas in pSIC they follow from Eq.\ \eqref{eq:forces_sic}. However, in the limit of an infinite system,  all three approaches would yield identical forces provided the energy depends quadratically on the polaron charge, an assumption made in our derivation of their equivalence (cf.\ Sec.\ \ref{sec:unified_hybrid} and Sec.\ \ref{sec:semilocal}).

% polaron density
We now compare the polaron densities obtained with $\gamma$DFT, $\mu$DFT, and  pSIC across all our case studies. For comparing with  pSIC, we consider the densities obtained for the neutral charge state ($q=0$) in the presence of polaronic lattice distortions. Indeed, within the assumptions of our semilocal formulation, variations in the polaron density lead to contributions to the total energy, which are higher than second order in the polaron charge $q$.  The polaron density is calculated as the square of the polaron wave function, which corresponds to the wave function of the highest occupied orbital in the spin up channel for electron polarons, or to the wave function of the lowest unoccupied orbital in the spin down channel for hole polarons. Then, we integrate the polaron densities over the $xy$ plane, namely
\begin{equation}
n_\text{p}(z) = \int dx dy \, n_\text{p}(x,y,z), \label{eq:average_density}
\end{equation}
where $n_\text{p}(x,y,z)$ is the polaron density defined on a three-dimensional grid. As illustrated in Fig.\ \ref{fig:densities}, we find excellent agreement between the results obtained with different piecewise-linear functionals. This further corroborates that piecewise-linear functionals are generally expected to lead to the same polaronic state. Exceptions to this general rule can occur when multiple polaronic states are in close energetic competition \cite{falletta2023PRB}. We note that the discrepancies in the polaronic structures obtained with the various functionals considered in this work are not found to have an effect on the overall shape and orientation of the polaron density. However, these properties might be affected by the use of a supercell of small size. For instance, in the case of the hole polaron in MgO, the polaron density obtained in the present study for the 216-atom supercell is oriented along the 001 direction. At variance, the polaron density obtained with PBE0($\alpha$), DFT+$U$, $\gamma$DFT, $\mu$DFT for the 64-atom supercell of MgO is oriented along the 111 direction \cite{kokott2018NJP,falletta2022PRL,falletta2022PRB,falletta2022npj}. In the latter case, the distorted O--Mg bonds are found to be all equivalent and equal to 2.23  \AA{} \cite{falletta2022PRB} with both $\gamma$DFT and $\mu$DFT \cite{falletta2024JAP}. In pSIC, the polaron fails to localize in the 64-atom MgO supercell. These discrepancies in the structure, consequent orientation of the polaron density, and localization can be attributed to finite-size effects, which could alter the relative stabilities of competing polaronic states.

% polaron formation energy
Next, we determine the $\gamma$DFT, $\mu$DFT, and  pSIC polaron formation energies for all polarons considered in this work. This is achieved by calculating the energy gain due to charge localization and the energy cost due to lattice distortions at the PBE level of theory for the obtained polaronic structures [cf.\ Eqs.\ \eqref{eq:E_PWL_semilocal} and \eqref{eq:Ef_2_sl}].  We report the formation energies in Table \ref{tab:Ef}. Overall, we find excellent agreement for the polaron formation energies obtained with $\gamma$DFT, $\mu$DFT, and pSIC. Specifically, discrepancies only amount to at most 0.13 eV, excluding the case of the $V_\text{k}$ center in NaI. This further corroborates the finding that polaron formation energies obtained with piecewise-linear functionals are robust irrespective of the specific functional adopted \cite{falletta2022PRL,falletta2022PRB,falletta2022npj,falletta2023PRB,falletta2024JAP}. This robustness is more likely guaranteed for materials with large dielectric constants (e.g., BiVO$_4$, TiO$_2$), as these would narrow the gap between charged and neutral polaron levels, thus suppressing finite-size effects on the polaron structure (cf.\ Fig.\ \ref{fig:finite-size}). 

Since formation energies are fully determined from the energetics of the neutral state at the PBE level of theory, differences between $\gamma$DFT, $\mu$DFT, and pSIC are only due to the use of different structures. As discussed previously, discrepancies in the structures  could also arise from the different treatment of finite-size effects. 
%AP: mi sembra di capire che le forze non sono uguali in questi tre schemi, quindi il problema finite-size è solo uno dei fattori che provoca una differenza. giusto? 
%SF: si esatto, le forze pSIC sono generalmente piu' deboli a causa dei finite size, quindi il pSIC ha piu' tendenza a localizzare. Invece le forze gDFT e mDFT sono piu' forti e hanno tendenza a localizzare.
For instance, in the cases of BiVO$_4$ and TiO$_2$, which are characterized by large high-frequency and static dielectric constants (cf.\ Table \ref{tab:dielectric_constants}), formation energies obtained with $\gamma$DFT, $\mu$DFT, and pSIC vary by at most 0.05 eV. At variance, in the case of the $V_\text{k}$ center in NaI, which has smaller dielectric constants (cf.\ Table \ref{tab:dielectric_constants}), we find larger discrepancies in formation energies, which relate to the large differences found for the I--I bond length (cf.\ Table \ref{tab:structures}). As the distortion increases, the cost of the polaronic distortions rises, leading to a destabilization of the polaron state.

\begin{table}[t]
    \caption{Formation energies of the various polarons considered in this work, obtained with $\gamma$DFT, $\mu$DFT, and  pSIC.}
	\label{tab:Ef}
	\begin{ruledtabular}
		\begin{tabular}{l *{3}{c} }	
		      System                                & $\gamma$DFT & $\mu$DFT & pSIC  \\ \hline \rule{-4pt}{3ex} \rule{-4pt}{3ex}
MgO ($h^+$)                           & $-0.31$     & $-0.27$  & $-0.40$                  \\ \rule{-4pt}{3ex}     
$\beta$-Ga$_2$O$_3$ ($h_\text{I}^+$)  & $-0.58$     & $-0.59$  & $-0.66$                  \\ \rule{-4pt}{3ex} 
NaI ($h^+$)                           & $-0.21$     & $-0.58$  & $-0.63$                  \\ \rule{-4pt}{3ex} 			 
BiVO$_4$ ($e^-$)                      & $-0.43$     & $-0.44$  & $-0.47$                  \\ \rule{-4pt}{3ex}   
TiO$_2$ ($e^-$)                       & $-0.38$     & $-0.36$  & $-0.41$                     \rule{-4pt}{3ex}
		\end{tabular}
	\end{ruledtabular}
\end{table}

% comparison with previous literature
% MgO: -0.20 (Varley), -0.38->-0.58 (Kokott, from a=0 to a=1), -0.53 (Falletta pbe0), -0.60 (Falletta DFT+U)
% Ga2O3: -0.11 (Varley), -0.62 eV (Deak), -0.58 (Gake), -0.54 (Frodason),  [ho2018PRB not found, Bouzid not found]], -0.63 (falletta hybrid), -0.74 (falletta DFT+U)
% NaI: -0.21 (sadigh) [miceli not found]
% BiVO4: -1.1 eV ()Wiktor),  -0.83 (pbe0 pham), -0.44 (falletta pbe0), -0.63 (falletta dft+u)
% TiO2: -0.14->-0.41 (kokott)

The present results for the polarons considered in this work can be compared with previous literature. The hole polaron in MgO was found to be stable using hybrid functionals \cite{varley2012PRB,kokott2018NJP,wing2020PRM,falletta2022PRB,falletta2022PRL} and the DFT+$U$ functional \cite{falletta2022npj}, with formation energies ranging from $-0.2$ eV to $-0.6$ eV. Dependencies of the formation energy on the fraction of Fock exchange are reported in Refs.\ \cite{kokott2018NJP,falletta2022PRL}. In contrast, the localization of the hole polaron in MgO was not found in Ref.\ \cite{wing2020PRM}. The hole polaron localized at the O$_\text{I}$ site in $\beta$-Ga$_2$O$_3$ was found using hybrid functionals \cite{varley2012PRB,deak2017PRB,gake2019PRM} and the DFT+$U$ functional \cite{falletta2023PRB}, with formation energies ranging from $-0.1$ to $-0.7$ eV. The $V_\text{k}$ center was found to be stable with the pSIC functional \cite{sadigh2015PRB} and hybrid functionals \cite{miceli2018PRB}. The formation energy obtained with pSIC in Ref.\ \cite{sadigh2015PRB} is $-0.27$ eV. This value does not account for finite-size corrections for the neutral defect with polaronic distortions. Upon inclusion of such corrections, the formation energy in Ref.\ \cite{sadigh2015PRB} amounts to $-0.48$ eV, deviating from the pSIC result obtained in this work by 0.15~eV. The electron polaron in BiVO$_4$ was found to be stable with hybrid functionals \cite{wiktorACS2018,pham2020JCTC,falletta2022PRL,falletta2022PRB} and DFT+$U$ functionals \cite{pham2020JCTC,falletta2022npj}, with formation energies ranging from $-0.4$ eV to $-1.1$ eV. In Ref.\ \cite{pham2020JCTC}, the formation energies are shown to depend noticeably on the fraction of Fock exchange and on the Hubbard $U$ parameter. The electron polaron in TiO$_2$ rutile was studied using hybrid functionals \cite{deskins2007PRB,pham2020JCTC} and DFT+$U$ functionals \cite{deskins2007PRB,pham2020JCTC,reticcioli2022JACM}, with substantial variations in formation energies reported for different Hubbard $U$ parameters \cite{pham2020JCTC}. Overall, the discrepancies in the formation energies obtained across all these previous works are due to differences in the functionals adopted (e.g., fraction of Fock exchange in hybrid functionals, value of $U$ in DFT+$U$), in the treatment of the finite-size corrections, in the supercell sizes used, and in the computational setups.

\vspace{0.7cm}
\subsection{Unit-cell method}

% coefficients A and B
We apply the unit-cell method of Sio \textit{et al.}\ \cite{sio2019PRL,sio2019PRB} to study the polarons considered in this work. The computational details for such calculations are given in Sec.\ \ref{sec:comput_details}. We start by performing an electronic ground state calculation, followed by a band calculation along the high-symmetry path and a phonon calculation. Next, we calculate the electron-phonon matrix elements in the Wannier representation of the relevant orbitals pertaining to valence-band or conduction-band states for hole or electron polarons, respectively, as outlined in Sec.\ \ref{sec:comput_details}. These electron-phonon matrix elements are then used to solve the self-consistent polaron equations \cite{sio2019PRL,sio2019PRB}. In this approach, the polaron wave function is expanded in terms of the single-particle Kohn-Sham states with coefficients $A_{n\mathbf k}$, where $n$ denotes the electronic band. The lattice distortions are expanded in terms of the phonon eigenmodes with coefficients $B_{\mathbf q\eta}$, where $\eta$ identifies the phonon branch. The  generalized Fourier amplitudes $A_{n\mathbf k}$ and $B_{\mathbf q\eta}$ obtained for the various polarons considered in this work are  given in the Supplemental Material (SM) \cite{SM}.

Having obtained the polaron wave function and the associated lattice distortions, the polaron formation energy is evaluated through Eq.\ \eqref{eq:Ef_sio}, where the absence of screening effects clearly appears. To account for the presence of finite-size effects, one needs to determine the polaron formation energy for increasingly dense $\mathbf k$-point meshes and extrapolate the result to the limit of an infinite supercell (dilute limit), as shown in the SM \cite{SM}. We find the extrapolated formation energies to be $-0.37$ eV for the hole polaron in MgO, $-0.73$ eV for the hole polaron in $\beta$-Ga$_2$O$_3$, $-0.40$ eV for the hole polaron in NaI, $-0.42$ eV for the electron polaron in BiVO$_4$, and $-0.46$ eV for the electron polaron in TiO$_2$. By inspecting the polaron wave functions, we find that the unit-cell method predicts the same polaronic states for MgO, BiVO$_4$, and TiO$_2$ shown in Fig.\ \ref{fig:polarons}. In these cases, the agreement for the formation energies obtained with the unit-cell method and pSIC is within 0.06 eV (Table \ref{tab:Ef}), further demonstrating the connection between these two approaches (Sec.\ \ref{sec:connection_pSIC_unitcell}).  For $\beta$-Ga$_2$O$_3$, the unit-cell method predicts the O$_{2\text{II}}$ hole polaron state, where the polaron is split between two O$_{\text{II}}$ sites. The formation energy for this state with the piecewise-linear hybrid functional was found to be $-0.71$ eV \cite{falletta2023PRB}, also demonstrating in this case the connection between piecewise-linear functionals and the unit-cell method, as described in Sec.\ \ref{sec:connection_pSIC_unitcell}. For NaI, the unit-cell method predicts a polaron state localized on a single I atom, rather than the V$_k$ center solution. These differences in polaron states encountered for $\beta$-Ga$_2$O$_3$ and NaI are due to different initializations of the polaronic distortions.

\section{Conclusions \label{sec:conclusions}}

% general summary of the results
In conclusion, we demonstrated the connection and equivalence of charged and neutral density functional formulations for addressing the many-body self-interaction of polarons. This was achieved by starting from the unified hybrid-functional formulation for the self-interaction and deriving a semilocal self-interaction-free functional. This derivation bridges the gap between various approaches for correcting the self-interaction of polarons, including the piecewise-linear semilocal functionals $\gamma$DFT and $\mu$DFT, the pSIC method, and the unit-cell method for polarons. We demonstrate that all these methods lead to the same formal expression of the self-interaction corrected energy, which can be determined using the semilocal energetics of the neutral defect with polaronic lattice distortions. The main differences between these methods result from the scheme leading to the polaron geometries, which are sensitive to the treatment of electron screening and finite-size effects. We apply these methods to a variety of electron and hole polarons, and demonstrate the agreement for the ground-state properties obtained with different piecewise-linear functionals. Weak polaronic distortions are found to be more sensitive to the functional adopted, and imply larger discrepancies in the polaron formation energies. At variance, in the case of materials with large dielectric constants, finite-size effects are less sizeable and results from piecewise-linear functionals are found in excellent agreement. 

% issue of finite size 
We emphasize that finite-size effects on the polaronic distortions are one of the main sources of discrepancies between the methods. In particular, $\gamma$DFT and $\mu$DFT are expected to overestimate the polaronic distortions, while pSIC is expected to underestimate the polaronic distortions. In this context, it would be suitable to employ a self-consistent finite-size approach to correct for finite-size effects when solving the Kohn-Sham equations. This is expected to lead to finite-size corrected wave functions and forces, which may narrow the discrepancies found for the polaron structures and formation energies obtained with $\gamma$DFT, $\mu$DFT, and pSIC. Furthermore, correcting for finite-size effects in the forces may prevent polaron delocalization induced by the use of  relatively small supercells. Efforts in this direction have been deployed  by da Silva \textit{et al}.\ \cite{silva2021PRL}, who used a self-consistent correction for the total energy. Further application to structural relaxations is however yet to be achieved and is left for future studies.

Overall, this work further highlights the advantages of using efficient semilocal functionals for predicting polaron properties without requiring large computational resources. Further comparisons between approaches for studying polarons are provided in recent works \cite{dai2025RMP,DAI2025JCP}. Molecular dynamics of polarons can be further accelerated by using machine-learning tools, such as equivariant neural networks \cite{batzner2023NC,musaelian2023NC,falletta2025NC}.

\section*{Acknowledgments}

The authors thank Kyle Bystrom, Feliciano Giustino, and Chuin Wei Tan for useful interactions. S.F.\ was supported by the Camille and Henry Dreyfus Foundation Grant No.\ ML-22-075 and by the Swiss National Science Foundation through the Postdoc mobility fellowship under grant number P500PT\_214445. This work was partially performed under the auspices of the US Department of Energy by Lawrence Livermore National Laboratory under contract DE-AC52-07NA27344. Computational resources were mainly provided by the Harvard University FAS Division of Science Research Computing Group. Calculations were also performed at the Swiss National Supercomputing Centre (CSCS) (grant under project No.\ s1122 and No.\ lp60) and at SCITAS-EPFL.

\newpage 

\bibliography{bibliography}

\end{document}